\documentclass[
reprint]{revtex4-2}
\usepackage{comment}
\usepackage{graphicx}
\usepackage{dcolumn}
\usepackage{bm}
\usepackage{xcolor}
\usepackage{cancel}
\newcommand{\ket}[1]{| {#1} \rangle} 
\newcommand{\bra}[1]{\langle {#1} |} 

\usepackage[utf8]{inputenc}
\usepackage[T1]{fontenc}
\usepackage[english,serbian]{babel}
\usepackage[OT1]{fontenc}
\usepackage{mathrsfs}
\usepackage{graphicx}
\usepackage{amssymb}
\usepackage{amsbsy}
\usepackage{latexsym}
\usepackage{mathtools}
\usepackage{titlesec}
\usepackage{amsmath}
\usepackage{amsthm}
\usepackage{color}
\usepackage[utf8]{inputenc}
\usepackage{lipsum}
\usepackage{fontenc}

\usepackage{chngcntr}

\def\nn{\nonumber}

\def\de{\textrm{d}}

\def\dj{d\kern-0.4em\char"16\kern-0.1em}
\def \Dj {\mbox{\raise0.3ex\hbox{-}\kern-0.4em D}}

\begin{document}

\preprint{APS/123-QED}

\title{Page Curve for an Evaporating Schwarzschild Black Hole in Dimensionally-Reduced Model of Dilaton Gravity}

\author{Stefan \Dj or\dj evi\'{c} and Voja Radovanovi\'{c}}
\affiliation{Faculty of Physics, University of Belgrade, Studentski Trg 12-16, 11000 Belgrade, Serbia}

\selectlanguage{english}

\date{\today}

\begin{abstract}

We study entanglement entropy of quantum fields in a (1+1)-dimensional model of dilaton gravity derived from the four-dimensional Einstein-Hilbert action by dimensional reduction. Scalar matter is coupled to gravity, while the back-reaction is included via the Polyakov-Liouville action. This theory exhibits both eternal and evaporating Schwarzschild black hole solutions. The fine-grained entropy in a collapse scenario is investigated by applying the "island rule" via the quantum extremal surface approach. We demonstrate that the fine-grained entropy of the Hawking radiation follows the Page curve, and therefore the evolution of the Hawking radiation is unitary.
\end{abstract}
\maketitle

\selectlanguage{english}
\section{Introduction}
\par The Hawking radiation and the information loss paradox \cite{Hawking1} are phenomena usually considered to capture the essential properties of quantum gravity and the microscopic nature of black hole entropy \cite{Bekenstein1, Bekenstein2, Hawking2}.  After gravitational collapse, the black hole starts to emit radiation known as Hawking radiation. This radiation is thermal in nature, implying that at the end of the evaporation the resulting state of radiation is a mixed state, even if the initial state of the collapsing matter was pure \cite{Hawking3}. This behavior breaks one of the fundamental principles of quantum mechanics - unitary time evolution, and is the essence of the information loss paradox. Hawking's calculation also demonstrated that the fine-grained entropy of the radiation continues to increase even beyond the Bekenstein-Hawking entropy limit, given by $S_{\text{BH}}=A(\text{horizon})/4G^{(4)}_{\text{N}}$ (we set $c=\hbar=k_{\text{B}}=1$, and $G^{(4)}_{\text{N}}$ stands for Newton's constant in four space-time dimensions). This is the other way to express the information loss paradox.
\par To investigate the information loss paradox, various toy models of (1 + 1) dimensional dilaton gravity, such as Jackiw-Teitelboim (JT) \cite{JT1,JT2} and Callan–Giddings–Harvey–Strominger (CGHS) \cite{CGHS}, were constructed. In these models, after integrating out quantum fluctuations of the matter fields and including $1$-loop quantum corrections, field equations can be exactly solved by introducing suitable correction terms as in Russo-Susskind-Thorlacius (RST), Bose-Parker-Peleg (BPP) and the CGSH model \cite{RST1,RST2,BPP,QCGHS}. Even when this is not the case, the two-dimensional models give less complicated equations, which in most cases can be solved perturbatively. Such is the case of the DREH (Dimensionally-Reduced Einstein-Hilbert) model \cite{DREH1, DREH2, Dimred}. A more comprehensive account of dilaton gravity models can be found in \cite{Fabri, DWV}.
\par The DREH model stands out due to its physical interpretation as a dimensionally-reduced version of general relativity. Classically, it is a two-dimensional model of dilaton gravity coupled to a scalar field. It admits the Schwarzschild black hole solution. In addition, gravitational collapse scenario can be constructed within this model. After adding the quantum corrections via the Polyakov-Liouville action, the eternal black hole solution was found in \cite{DREH1}, as well as an evaporating black hole solution \cite{DREH2}.
\par In the nineties, Page suggested that the fine-grained entropy should follow a certain curve \cite{Page_1993,Page_2013}. At the beginning, the entropy should increase part, as Hawking's result suggests. However, at late times the fine-grained entropy should decrease and ultimately vanish at the end-point of the evaporation. For many two-dimensional dilaton gravity models, the Page curve could not be reproduced. The final conclusion was that the information is lost \cite{RST1,RST2,QCGHS,BPP}. 
\par At present, with the help of a gravitational path integral and QES prescription \cite{Takayanagi1,Takayanagi2,QES} a new formula is derived for the entanglement entropy in gravitational systems, the so-called "island formula" \cite{penington,islands6,islands5,PagecurveMaldecena,MaldecenaJT}:
\begin{equation}\label{Island_formula}
S_{FG}(R)=\underset{I}{\min}\left\{\underset{I}{\text{ext}}\left[\frac{\text{A}(\partial I)}{4G_{N}}+S_{\text{matter}}(R\cup I)\right]\right\}.
\end{equation}
\par Namely, it turns out that there are two saddle points to the gravitational path integral, contrary to previous opinion that there is only one, the Hawking's saddle. Using the replica-wormholes method, another saddle point was found, the replica-wormholes saddle. The Hawking's saddle gives an ever-increasing contribution to the fine-grained entropy, while the replica-wormhole saddle results in a monotonically decreasing contribution. Taking the minimum of these two contributions, the formula (\ref{island}) successfully reproduces the Page curve. Since the replica wormhole method is merely a consequence of the replica trick in models with gravitation, the island rule is expected to be applicable to any kind of black hole.
\par The Page curve has so far been reproduced in a variety of two-dimensional dilaton gravity models \cite{penington1,islands4,islands5,islands6,islands7,islands8,islands9,islands10,islands11,islands12,islands13,islands14,MaldecenaJT,verlinde,RSTislands,BPPislands,notes,DREH1}, but also in higher-dimensional models \cite{islands1,islands2,Schislands,islands3,islands_RN,Islands_Kerr}. Some other interesting studies regarding entanglement entropy in gravitational systems can be found in \cite{EE1,EE2,EE3,EE4,EE5}.
\par The key outcome of this paper is the reproduction of the Page curve for the evaporating Schwarzschild black hole. Additionally, the location of the QES (or the island) is notably positioned behind the event horizon of the black hole, as expected. All computations are performed within a perturbative framework, extending to  first order in Planck's constant.
\par This paper is organized as follows. In the subsequent section, we give a quick review of a (1+1)-dimensional model of dilaton gravity derived from the Einstein-Hilbert action by dimensional reduction. In addition, we describe classical and quantum-corrected collapse scenarios within this model \cite{DREH2}. Section III investigates the fine-grained entropy of the Hawking radiation in the classical collapse scenario, and results in an incorrect Page curve since the back-reaction has not been included. The correct Page curve, within the quantum-corrected collapse scenario, is computed in section IV, up to first order in $\hbar$. Both sections III and IV consist of two parts. In the first parts, we consider the contribution to the fine-grained entropy of the radiation that comes from the Hawking's saddle point, while in the second parts the replica-wormholes saddle point's contribution is studied. The brief summery and conclusion with some proposals for the future work are given in Section V.    
\section{DREH model and the evaporating black hole solution}
\par In this section we will discuss the evaporating black hole solution of the DREH model. The DREH model is the (1+1)-dimensional model of dilaton gravity obtained from the usual four-dimensional Einstein-Hilbert (EH) action by using the spherically symmetric ansatz and integrating out the angles degrees of freedom. The model is, in some ways, similar to the CGHS model of dilaton gravity \cite{CGHS}. In particular, it admits black hole solutions. 
\par Dimensional reduction is a well-known procedure. The technical details can be found in \cite{MVM} and \cite{DREH1,DREH2}. Starting with $4D$ Einstein-Hilbert action, and preforming dimensional reduction, we arrive at the following two-dimensional dilaton gravity theory:  
\begin{equation}\label{Dilaton_action}
S_{\phi}=\frac{1}{4G}\int \de^{2}x\sqrt{-g}\Big[e^{-2\phi}\left(R+2(\nabla\phi)^{2}\right)+2\lambda^{2}\Big],    
\end{equation}
where we introduced $G\equiv\lambda^{2}G_{\text{N}}^{(4)}$ as Newton's constant of the reduced theory; $\phi$ is a dilaton field, which represents an artifact of the dimensional reduction procedure, while $\lambda$ is a dimensional constant introduced for the purpose of dimensional reduction.
\par Next we introduce a conformal matter term $S_{m}$, as massless scalar field $f$ minimally coupled to gravity, which results in the DREH action:
\begin{align}
&S_{\text{DREH}}=S_{\phi}+S_{m}\nn\\
&=\frac{1}{4G}\int \de^{2}x\sqrt{-g}\Big[e^{-2\phi}\left(R+2(\nabla\phi)^{2}\right)+2\lambda^{2}\Big]\nn\\
&-\frac{1}{2}\int \de^{2}x\sqrt{-g}\left(\nabla f\right)^{2}.\label{dejstvo}
\end{align}
After varying the action (\ref{dejstvo}), one arrives at the classical equations of motion. For the metric, we use the conformal gauge $\mathrm{d}s^{2}=-e^{2\rho}\mathrm{d}x^{+}\mathrm{d}x^{-}$. The general solution to the vacuum equations of motion can be expressed as $(\mathrm{F}^{+},\mathrm{F}^{-},a)$, where $\mathrm{F}^{\pm}=\mathrm{F}^{\pm}(x^{\pm})$ are arbitrary functions, while $a$ is an arbitrary constant. The solution is given by (for more details see \cite{DREH2}):
\begin{align}
    \varphi+\lambda a\ln{\left(\frac{\varphi}{\lambda a}-1\right)}&=-\frac{\lambda}{2}\left(\int\frac{\de x^{+}}{\mathrm{F}^{+}}+\int\frac{\de x^{-}}{\mathrm{F}^{-}}\right),\label{chi_resenje}\\
    \mathrm{F}^{+}\mathrm{F}^{-}e^{2\rho}&=\frac{\lambda a}{\varphi}-1,\label{rho_resenje}
\end{align}
where we introduced another field $\varphi=e^{-\phi}$. It is easy to show that $\rm{F}^{\pm}$ represent coordinate transformations that transform $x^{\pm}$ coordinates into the Eddington-Finkelstein coordinates $\sigma^{\pm}$. This means that the solution given by $(-1,1,a)$ corresponds to the Eddington-Finkelstein gauge. The constant $a=\frac{2MG}{\lambda^{2}}$ is a Schwarzschild radius. All of this implies that the solution (\ref{chi_resenje}-\ref{rho_resenje}) is a classical Schwarzschild black hole solution. Equation (\ref{chi_resenje}) represents the definition of a tortoise coordinate, while equation (\ref{rho_resenje}) gives us the form of the metric. To simplify the equations, we usher in a reduced field: $\mathrm{x}=\frac{\varphi}{\lambda a}$.
\par To explore gravitational collapse, we introduce matter through the energy-momentum tensor (EMT) of the form $T_{++}=M\delta(\sigma^{+}-\sigma^{+}_{0})$. This definition of EMT corresponds to the infalling light-like matter along the hypersurface given by $\sigma^{+}=\sigma^{+}_{0}$. Continuously connecting two vacuum solutions $(-1,1,0)$ (Minkowski vacuum) and $(\mathrm{F}^{+},\mathrm{F}^{-},a)$, along the $\sigma^{+}=\sigma^{+}_{0}$ hypersurface, one arrives at the following solution for a black hole formed in the gravitational collapse:
\begin{align}
        \mathrm{x}&=-\ln{\delta}+\hat{\mathrm{x}}=\frac{r_{*}}{a},\hspace{10mm}\sigma^{+}\leqslant \sigma^{+}_{0},\label{colaps_<_x_resenje}\\
        \mathrm{x}+\ln{\left(\mathrm{x}-1\right)}&=-\ln{\delta}+\hat{\mathrm{x}}+\ln{\left(\hat{\mathrm{x}}-1\right)}=\frac{r_{*}}{a},\hspace{0mm}\sigma^{+}>\sigma^{+}_{0},\label{colaps_>_x_resenje}
\end{align}
\begin{equation}
        \mathrm{F}^{+}\mathrm{F}^{-}e^{2\rho}=\begin{cases}
            -1,&\sigma^{+}\leqslant \sigma^{+}_{0}\\
            \frac{1}{\mathrm{x}}-1,&\sigma^{+}>\sigma^{+}_{0}
        \end{cases},\label{kolaps_metrika_resenje}
\end{equation}
where we have introduced a new set of coordinates $\hat{\mathrm{x}}(\sigma^{-})=\frac{\sigma^{+}_{0}-\sigma^{-}}{2a}$ and $\delta(\sigma^{+})=\exp{\left(-\frac{\sigma^{+}-\sigma^{+}_{0}}{2a}\right)}$, as well as the tortoise coordinate $r_{*}$. The functions $\mathrm{F}^{\pm}$ are given by:
\begin{equation}
    \mathrm{F}^{+}=-1\hspace{2mm}\text{and}\hspace{2mm}\mathrm{F}^{-}=\frac{\hat{\mathrm{x}}-1}{\hat{\mathrm{x}}}.\label{FpFm}
\end{equation}
Note that the asymptotically flat coordinates change from $\sigma^{\pm}$ to $\hat{\sigma}^{\pm}$ defined by the functions $\mathrm{F}^{\pm}$. After the integration, they are given by:
\begin{align}
    \hat{\sigma}^{+}&=\sigma^{+},\label{sigma_hat_+}\\
    \hat{\sigma}^{-}&=\sigma^{+}_{0}-2a\bigg{(}\hat{\mathrm{x}}+\ln{\left|\hat{\mathrm{x}}-1\right|}\bigg{)}.\label{sigma_hat_-}
\end{align}
The black hole resulting from this process features an event horizon and a singularity, each described by specific hypersurfaces, respectively, given by:
\begin{align}
    &\sigma^{-}_{\mathrm{H}}=\sigma^{+}_{0}-2a\label{horizont},\\
    &\hat{\mathrm{x}}_{\mathrm{Sing}}+\ln{\left(\hat{\mathrm{x}}_{\mathrm{Sing}}-1\right)}-\ln{\delta_{\mathrm{Sing}}}=0.\label{singularnost}
\end{align}
\par Since the goal is to investigate the process of black hole evaporation, we need to add quantum corrections, by quantizing the scalar field $f$. These quantum corrections come in the form of the Polyakov-Liouville (PL) action \cite{Polyakov}:

\begin{align}\label{SPL}
S_{\text{PL}}=-\frac{\hbar}{96\pi}\int\de^{2}x\int&\de^{2}x'\sqrt{-g(x)}\sqrt{-g(x')}\nonumber \\
&\times R(x)G(x-x')R(x'),
\end{align}
After the localization of the action (\ref{SPL}) and by subsequent variation of this action, one derives the equations of motion of the quantum-corrected theory. These equations cannot be solved exactly, but can be solved perturbatively with respect to $\varepsilon=\frac{\hbar G}{12\pi}$. The general solution can be expressed as $(\mathrm{F}^{+},\mathrm{F}^{-},a,t_{+},t_{-})$, where $t_{\pm}(x^{\pm})$ are functions that define the quantum state of the radiation, and represent the normally ordered part of the energy-momentum tensor of quantum fields. To establish the evaporation black hole scenario, space-time has to exhibit an energy flux at the future null-infinity, while there is no energy flux at past null-infinity. This means that the EMT must be normally ordered in Minkowski coordinates $\sigma^{\pm}$, which corresponds to the choice $t_{\pm}(\sigma^{\pm})=0$. 
\par
\begin{figure}
    \begin{center}
    \includegraphics[width=7cm, height=7cm]{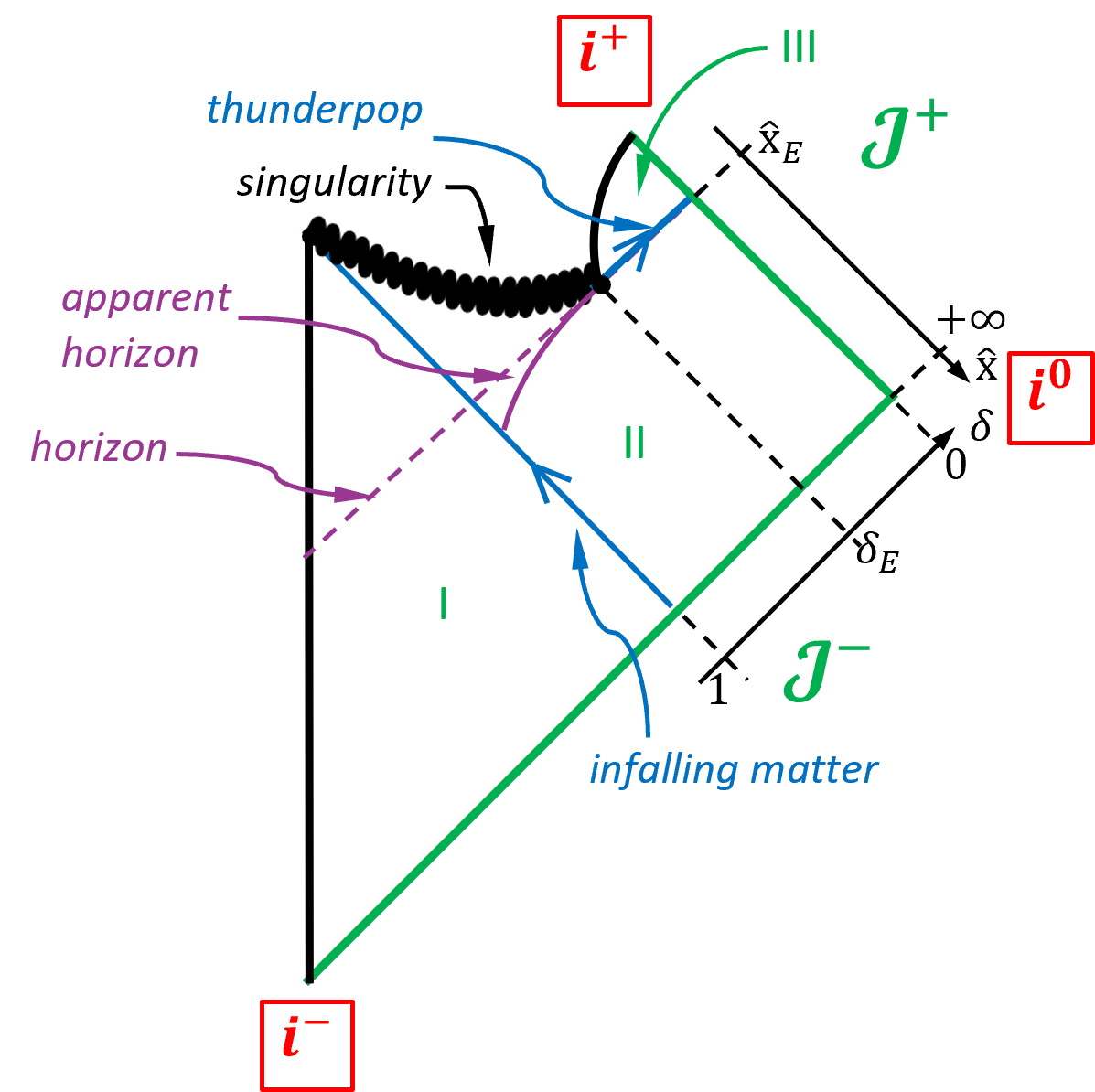}
    \caption{This Penrose diagram shows space-time of an evaporating black hole. Three distinct regions of space-time, as well as the $\delta$-axis and $\hat{\mathrm{x}}$-axis, are explicitly shown in the figure. The event horizon is represented by the dashed purple line, while the apparent horizon is depicted by the solid purple line. The shockwaves are shown in blue, and the singularity is colored black as is the boundary of space-time. The coordinates of the end-point of the evaporation $(\delta_{E},\hat{\mathrm{x}}_{E})$ are presented along the corresponding axis.}\label{sl1}
    \end{center}
\end{figure}

 To better understand how space-time looks like, it is instructive to give a conformal diagram, Figure \ref{sl1}. There exist three distinctive parts of space-time: part I corresponds to the Minkowski space-time that existed before the creation of the black hole; part II is the part that exhibits the evaporating black hole; while part III is what is left of space-time after the black hole evaporates completely.
\par After some very complex calculations (see \cite{DREH2}), one arrives at the following solution for an evaporating black hole, in the region II of space-time where the black hole exists:
\begin{widetext}
\begin{align}
    &\mathrm{F}^{+}\mathrm{F}^{-}e^{2\rho}=\left(\frac{1}{\mathrm{x}}-1\right)\left\{1+\frac{\varepsilon}{8(\lambda a)^{2}}\left[\frac{\mathrm{x}-1}{\mathrm{x}}\frac{\de\mathrm{S}^{>}_{0}(\mathrm{x})}{\de\mathrm{x}}-\frac{2\ln{\delta}}{\mathrm{x}-1}-\sum_{n=1}^{\infty}\frac{(1-\delta)^{n}}{n!}\left(\frac{\mathrm{x}-1}{\mathrm{x}}\frac{\de\mathrm{S}^{>}_{n}(\mathrm{x})}{\de\mathrm{x}}+\mathrm{S}^{>}_{n+1}(\mathrm{x})-n\mathrm{S}^{>}_{n}(\mathrm{x})\right)\right]\right\}\nonumber\\
    &\hspace{7cm}\times\left[1-\frac{\varepsilon}{8(\lambda a)^{2}}\sum_{n=1}^{\infty}\frac{(1-\delta)^{n}}{n!}(z_{n+1}-nz_{n})\right]\label{rho_resenje_>}\\
    &\mathrm{F}^{+}\mathrm{F}^{-}e^{2\rho}=\left(\frac{1}{\mathrm{x}}-1\right)\left\{1+\frac{\varepsilon}{8(\lambda a)^{2}}\left[\frac{\mathrm{x}-1}{\mathrm{x}}\frac{\de\mathrm{S}^{<}_{0}(\mathrm{x})}{\de\mathrm{x}}+\mathrm{S}^{>}_{1}(\mathrm{x})-\frac{2\ln{\delta}}{\mathrm{x}-1}-\sum_{n=1}^{\infty}\frac{\delta^{n}}{n!}\left(\frac{\mathrm{x}-1}{\mathrm{x}}\frac{\de\mathrm{S}^{<}_{n}(\mathrm{x})}{\de\mathrm{x}}-n\mathrm{S}^{<}_{n}(\mathrm{x})\right)\right]\right\}\nonumber\\
    &\hspace{7cm}\times\left[1-\frac{\varepsilon}{8(\lambda a)^{2}}\sum_{n=1}^{\infty}\frac{(1-\delta)^{n}}{n!}(z_{n+1}-nz_{n})\right]\label{rho_resenje_<}\\
    &\mathrm{x}+\ln{(\mathrm{x}-1)}-\frac{\varepsilon}{8(\lambda a)^{2}}\left[\mathrm{S}_{0}^{>}(\mathrm{x})-\frac{\pi^{2}}{3}-2+\ln{\delta}\mathrm{S}_{-1}(\mathrm{x})-\sum_{n=1}^{\infty}\frac{(1-\delta)^{n}}{n!}(\mathrm{S}_{n}^{>}(\mathrm{x})+z_{n})\right]\nonumber\\
    &\hspace{7cm}=-\ln{\hat{\delta}}+\hat{\mathrm{x}}+\ln{(\hat{\mathrm{x}}-1)}-\frac{\varepsilon}{8(\lambda a)^{2}}\mathrm{S}^{>}_{0}(\hat{\mathrm{x}})\equiv\frac{\hat{\sigma}^{+}-\hat{\sigma}^{-}}{2a}\equiv\frac{\hat{\sigma}}{a},\label{x_resenje_>}\\
    &\mathrm{x}+\ln{(1-\mathrm{x})}-\frac{\varepsilon}{8(\lambda a)^{2}}\left[\mathrm{S}_{0}^{<}(\mathrm{x})-\frac{\pi^{2}}{3}-2+\ln{\delta}\mathrm{S}_{-1}(\mathrm{x})-\sum_{n=1}^{\infty}\frac{\delta^{n}}{n!}\mathrm{S}_{n}^{<}(\mathrm{x})-\sum_{n=1}^{\infty}\frac{(1-\delta)^{n}}{n!}z_{n}\right]\nonumber\\
    &\hspace{7cm}=-\ln{\hat{\delta}}+\hat{\mathrm{x}}+\ln{(1-\hat{\mathrm{x}})}-\frac{\varepsilon}{8(\lambda a)^{2}}\mathrm{S}^{>}_{0}(\hat{\mathrm{x}})\equiv\frac{\hat{\sigma}^{+}-\hat{\sigma}^{-}}{2a}\equiv\frac{\hat{\sigma}}{a}.\label{x_resenje_<}
\end{align}
\end{widetext}
\noindent Equation (\ref{rho_resenje_>}) represents the solution for the metric for $\mathrm{x}\geqslant1$, while equation (\ref{rho_resenje_<}) is the solution for the metric for $\mathrm{x}\leqslant1$. Equations (\ref{x_resenje_>}) and (\ref{x_resenje_<}) are the equations for the tortoise coordinate for $\mathrm{x}\geqslant1$ and $\mathrm{x}\leqslant1$, respectively, in terms of coordinates $\sigma^{+}$ (through $\delta$ dependence) and $\sigma^{-}$ (through dependence $\hat{\mathrm{x}}$. When $\delta=1$ both (\ref{rho_resenje_>}) and (\ref{rho_resenje_<}) reduce to equation for the coordinate transformation $\mathrm{F}^{-}$. This implies that $e^{2\rho}=1$ when $\delta=1$, which is the continuity condition for the metric. Also, equations (\ref{x_resenje_>}) and (\ref{x_resenje_<}) imply that $\mathrm{x}=\hat{\mathrm{x}}$ when $\delta=1$, which is the continuity condition for the dilaton field. In equations (\ref{x_resenje_>}) and (\ref{x_resenje_<}) we can see new asymptotically flat coordinates $\hat{\sigma}^{\pm}$ and $\hat{\delta}=\exp{\left(-\frac{\hat{\sigma}^{+}-\sigma^{+}_{0}}{2a}\right)}$. For more information on functions $S_{k}^{>/<}(\mathrm{x})$ and constants $z_{n}$ appearing in equations (\ref{rho_resenje_>}-\ref{x_resenje_>}) see Appendix \ref{app_1} or \cite{DREH2}.
\par To see what happens in region III of space-time, one needs to find the end-point of the evaporation. This point is defined as the point when the singularity hypersurface intersects with the apparent horizon and the event horizon. This intersection can be found, but one has to be careful since the perturbative approach breaks at this point. The end-point is given by \cite{DREH2}:
\begin{align}
     \delta_{E}&=e^{-\frac{4(\lambda a)^{2}}{\varepsilon}\left(1-\frac{\sqrt{\varepsilon}}{\lambda a}\right)},\label{ln_delta_E_razvoj}\\
     \hat{\mathrm{x}}_{E}&=1+\frac{\varepsilon}{4(\lambda a)^{2}}\left[1+e^{-\frac{4(\lambda a)^{2}}{\varepsilon}\left(1-\frac{\sqrt{\varepsilon}}{\lambda a}\right)}\right].\label{X_hat_E_razov_resenje}
\end{align}
The event horizon is defined as $\hat{\mathrm{x}}_{H}=\hat{\mathrm{x}}_{E}$, using (\ref{x_resenje_<}) it is possible to find the horizon as a hypersurface $\mathrm{x}(\delta)$ \cite{DREH2}:
\begin{equation}
    \mathrm{x}_{H}(\delta)=\left(1+\frac{\varepsilon}{4(\lambda a)^{2}}\ln{\delta}\right)\left[1+\frac{\varepsilon}{4(\lambda a)^{2}}\left(1+\frac{\delta_{E}}{\delta}\right)\right].\label{x_H}
\end{equation}
Another important quantity is the apparent horizon. It can be defined in terms of $\mathrm{x}=\mathrm{x}(\delta)$:
\begin{equation}
    \mathrm{x}_{AH}(\delta)=1+\frac{\varepsilon}{4(\lambda a)^{2}}\bigg{(}2+\ln{\delta}\bigg{)};\label{AH}
\end{equation}
or in terms of $\hat{\mathrm{x}}=\hat{\mathrm{x}}(\delta)$:
\begin{equation}
    \hat{\mathrm{x}}_{AH}(\delta)=1+\frac{\varepsilon}{4(\lambda a)^{2}}(1+\delta).\label{x_hat_AH_early}
\end{equation}
The singularity is expressed as:
\begin{equation}
    \mathrm{x}_{S}=\frac{\sqrt{\varepsilon}}{\lambda a}.\label{Singularity}
\end{equation}
After using equations (\ref{ln_delta_E_razvoj}), (\ref{X_hat_E_razov_resenje}) and reintroducing the physical constants, one arrives at the following result for the time of evaporation of the black hole:
\begin{equation}
    \Delta t_{E}=\frac{384M^{3}G^{2}}{\hbar\lambda^{4}}\bigg{(}1+\mathcal{O}(\sqrt{\varepsilon})\bigg{)}.
\end{equation}
It is important to note that the dominant term behaves as $M^{3}$,  which is the expected result from a thermodynamical standpoint, as well as the result derived in \cite{Dimred}. It is straightforward to show that up to the end-point of the evaporation, the black hole has evaporated almost all of its mass; the remainder being of $\mathcal{O}(\sqrt{\varepsilon})$ order. In \cite{DREH2} we showed that the end-state geometry of region III is, once again, Minkowski space-time. There is also a thunderpop at the end-point of evaporation, as in many other models of two-dimensional dilaton gravity \cite{RST1,RST2,BPP}.
\section{Page curve within the classical collapse scenario}

In this section, we analyze the Page curve for the evaporating black hole in the classical DREH model. Even though we cannot determinate what happens at the end-point of the evaporation (since it is located at future time infinity within the classical limit), we can still get some insight into the behavior of the Page curve in zeroth order with respect to $\varepsilon$. The time dependence of the fine-grained entropy of the Hawking radiation is given by the QES formula (\ref{Island_formula}). It comes down to the process of extremization of the generalized entropy formula:
\begin{equation}
    S_{gen}=\frac{A(\partial I)}{4G}+S_{\mathrm{matter}},\label{s_gen}
\end{equation}
where the first term is an area term and the second term represents the entropy of the quantum fields. Essentially, there exist two saddle points. One saddle point is located at the boundary of space-time, and it gives the ever-increasing contribution to the fine-grained entropy, which reproduces famous Hawking's result. For that reason, this saddle point is called Hawking's saddle point. Since it is located at the boundary of space-time, the area term in (\ref{s_gen}) vanishes. As time passes more particles leave the event horizon, which leads to the rise in the matter term. 
\par The other saddle point is located at the boundary of another region called the island $I$, and gives a decreasing contribution to the entropy. The boundary of the island region lies close to the event horizon. For that reason, the matter term is much smaller than the area term. With the passage of time, the event horizon shrinks and, in turn, the area term in (\ref{s_gen}) decreases. At the beginning of the evaporation, the dominant saddle point is the Hawking's saddle point. At a certain point in time, known as Page time, there occurs a transition between the two saddle points. After Page time, the dominant contribution to the fine-grained entropy comes from the so-called the replica-wormhole saddle point (named for the method by which it has been derived). We will consider two separate cases: one for which $I=\emptyset$ all the time (no island), and the other when the island appears at some point.
\par We assume that a black hole can be regarded as a simple quantum system up to a specific time-like hypersurface, known as the cut-off surface. In Figures \ref{sl2} and \ref{sl3} it is represented by a purple line.
\subsection{No-island case}\label{Sec_no_island_class}

\begin{figure}
    \begin{center}
    \includegraphics[width=6cm, height=7cm]{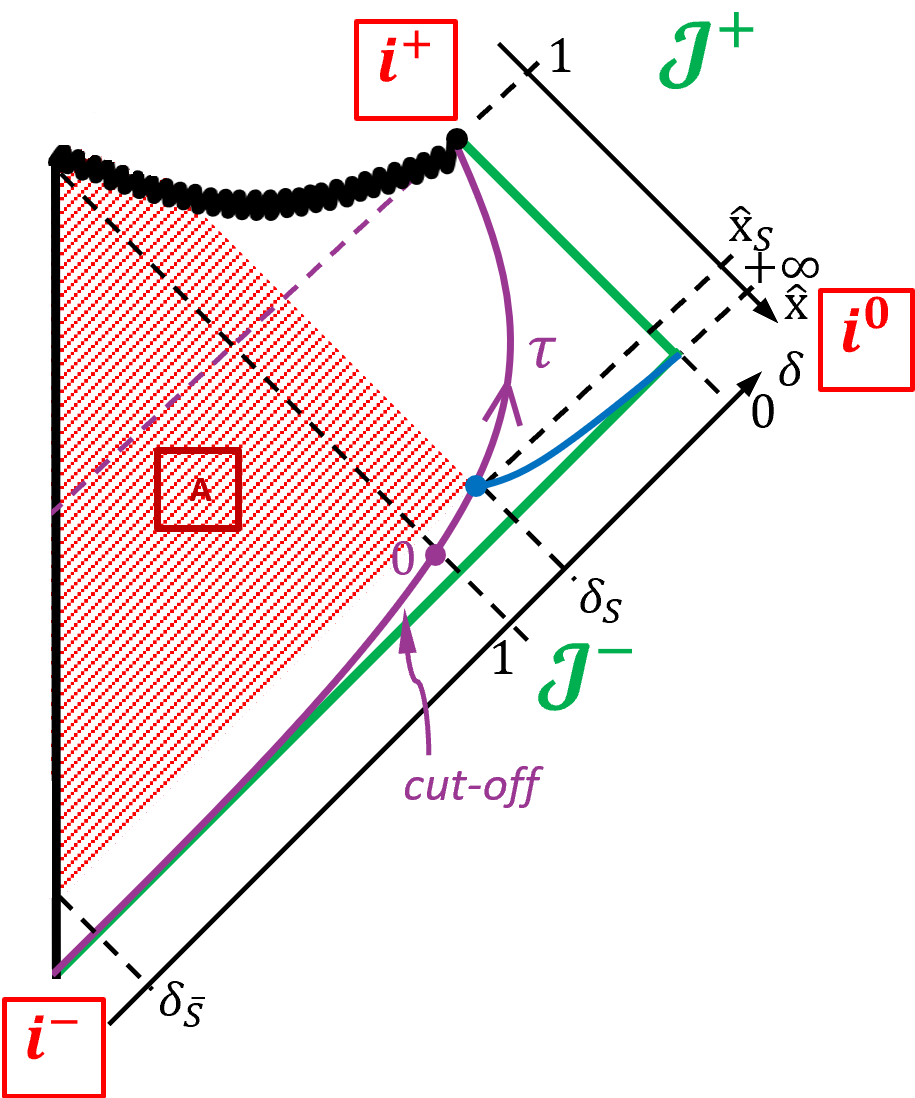}
    \caption{Position of the region ($A$) in which the inaccessible degrees of freedom live for a no-island classical case is shown in red. The cut-off hypersurface is depicted in purple. There is also a $\tau$-axis along the cut-off surface. The degrees of freedom of the radiation live in a  causal diamond over a constant-time slice shown in blue. Coordinates of all relevant points ($\delta_{S}$, $\hat{\mathrm{x}}_{S}$), as well as the ones obtained by the reflection from the boundary of space-time ($\delta_{\bar{S}}$), are presented on both $\delta$-axis and $\hat{\mathrm{x}}$-axis. Note that the region III does not exist since the quantum-corrections have not been included.}\label{sl2}
    \end{center}
\end{figure}

Initially, we examine the scenario in the absence of the island. In this case, we expect to reproduce Hawking's original prediction. The generalized entropy (\ref{s_gen}) reduces to the matter term. Since the vacuum is defined in the asymptotically flat coordinates of region I of space-time, the entanglement entropy should be expressed in these coordinates as well. Since there is a boundary within space-time, the correct formula for the entropy of quantum fields is (\ref{ent_no_island}), i.e.
\begin{equation}
    S_{\mathrm{matter}}=\frac{1}{12}\ln{\frac{(\sigma_{S}^{+}-\sigma_{\bar{S}}^{+})^{2}}{4\epsilon^{2}e^{-2\rho_{S}(\sigma)}}}.\label{no_island}
\end{equation}
The label $S$ corresponds to the boundary of the black hole region, as shown in Figure \ref{sl2}. We work in $(t,r_{*})$ coordinates. Then, the coordinates on the cut-off surface are $(t,b_{*})$, or $(t,b)$, where $\frac{b}{a}+\ln{\left(\frac{b}{a}-1\right)}=\frac{b_{*}}{a}$. It is important to note that time $t$ is defined with respect to the asymptotically flat coordinates of region II (\ref{sigma_hat_+}-\ref{sigma_hat_-}), i.e. $t=\frac{\hat{\sigma}^{+}-\hat{\sigma}^{-}}{2}$. The cut-off surface should be placed near asymptotic infinity, which implies that $b_{*}$ is very large. In the remainder of this paper we assume that $\frac{b_{*}}{a}\gg\frac{(\lambda a)^{2}}{\varepsilon}$. The parameter $\epsilon$ is a UV cut-off.    
\par The reflective boundary conditions at the $\mathrm{x}=0$ boundary imply that $\sigma^{+}_{\bar{S}}=\sigma^{-}_{S}$. The initial time of evaporation is defined by the intersection of the hypersurfaces $\sigma^{+}=\sigma^{+}_{0}$ and $r_{*}=b_{*}$, that is, $t_{0}=\sigma^{+}_{0}-b_{*}$. Then, the time of evaporation is defined as $\tau=t-t_{0}$. Using equation (\ref{colaps_>_x_resenje}), the cut-off hypersurface in region II of space-time is given by:
\begin{equation}
    \frac{b_{*}}{a}=-\ln{\delta_{S}}+\hat{\mathrm{x}}_{S}+\ln{(\hat{\mathrm{x}}_{S}-1)}.\label{cutoff_class}
\end{equation}
Note that the cut-off surface belongs to the region outside of the event horizon of the black hole, defined by equation (\ref{horizont}), which implies that $\hat{\mathrm{x}}_{S}>1$. With the help of equations (\ref{sigma_hat_+}) and (\ref{sigma_hat_-}), it is easy to see that the time of evaporation $\tau$ is connected to $\hat{\mathrm{x}}_{S}$ through the following transcendental equation:
\begin{equation}
    \hat{\mathrm{x}}_{S}+\ln{(\hat{\mathrm{x}}_{S}-1)}=\frac{b_{*}}{a}-\frac{\tau}{2a}.\label{veza_tau_x_hat_S}
\end{equation}
We demand $S_{FG}(0)=0$, which regularizes the UV divergencies in the expression for the entanglement entropy (\ref{no_island}). In terms of $\hat{\mathrm{x}}_{S}$, the fine-grained entropy along the Hawking's saddle point is given by:
\begin{equation}
    S_{H}(\hat{\mathrm{x}}_{S})=\frac{1}{12}\ln{\frac{a^{2}(b-a)\hat{\mathrm{x}}_{S}\left(\frac{b_{*}}{a}-\ln{(\hat{\mathrm{x}}_{S}-1)}\right)^{2}}{b^{3}(\hat{\mathrm{x}}_{S}-1)}}.\label{S_FG_no_island_class}
\end{equation}
The explicit dependence of the fine-grained entropy (\ref{S_FG_no_island_class}) on the evaporation time $\tau$ cannot be analytically obtained since one needs to solve the transcendental equation (\ref{veza_tau_x_hat_S}) that connects time $\tau$ and coordinate $\hat{\mathrm{x}}_{S}$. On the other hand, one can derive the expressions for fine-grained entropy at the beginning of the evaporation as well as at late times. At the beginning of the evaporation ($\tau\to0$) equation (\ref{veza_tau_x_hat_S}) can be perturbatively solved:
\begin{equation}
    \hat{\mathrm{x}}_{S}^{(0)}=\frac{b}{a}-\frac{b-a}{2ab}\tau.\label{a1}
\end{equation}
Substituting the result (\ref{a1}) into equation (\ref{S_FG_no_island_class}) and expanding with respect to $\tau$, one gets the following expression for the fine-grained entropy at early times:
\begin{equation}
    S_{H}^{(0)}(\tau)=\frac{a\tau}{8b^{2}}.\label{S_FG_early_times_class}
\end{equation}
At late times, quantity $\exp{\left(\frac{b_{*}}{a}-\frac{\tau}{2a}\right)}$ becomes very small, and so we can expand $\hat{\mathrm{x}}_{S}$ in terms of this quantity:
\begin{equation}
    \hat{\mathrm{x}}_{S}^{(\infty)}=1+e^{\frac{b_{*}}{a}-1}e^{-\frac{\tau}{2a}}.\label{a2}
\end{equation}
Substituting this equation (\ref{a2}) into equation (\ref{S_FG_no_island_class}) we get the expected Hawking's result of linear growth of the fine-grained entropy at late times:
\begin{equation}
    S_{H}^{(\infty)}(\tau)=\frac{\tau}{24a}+\frac{1}{12}\ln{\frac{(b-a)\tau^{2}}{4b^{3}}}-\frac{1}{12}\left(\frac{b_{*}}{a}-1\right).\label{S_FG_late_times_class}
\end{equation}
Notice that at late times, the dominant term in (\ref{S_FG_late_times_class}) is of type $\frac{\kappa\tau}{12}$, where $\kappa=\frac{1}{2a}$ is the standard surface gravity of a Schwarzschild black hole. This result is in agreement with the calculation of the fine-grained entropy in many models of $2D$ dilaton gravity \cite{BPPislands,RSTislands}.
\subsection{Island case}\label{Sec_island_class}
\begin{figure}
    \begin{center}
    \includegraphics[width=6cm, height=7cm]{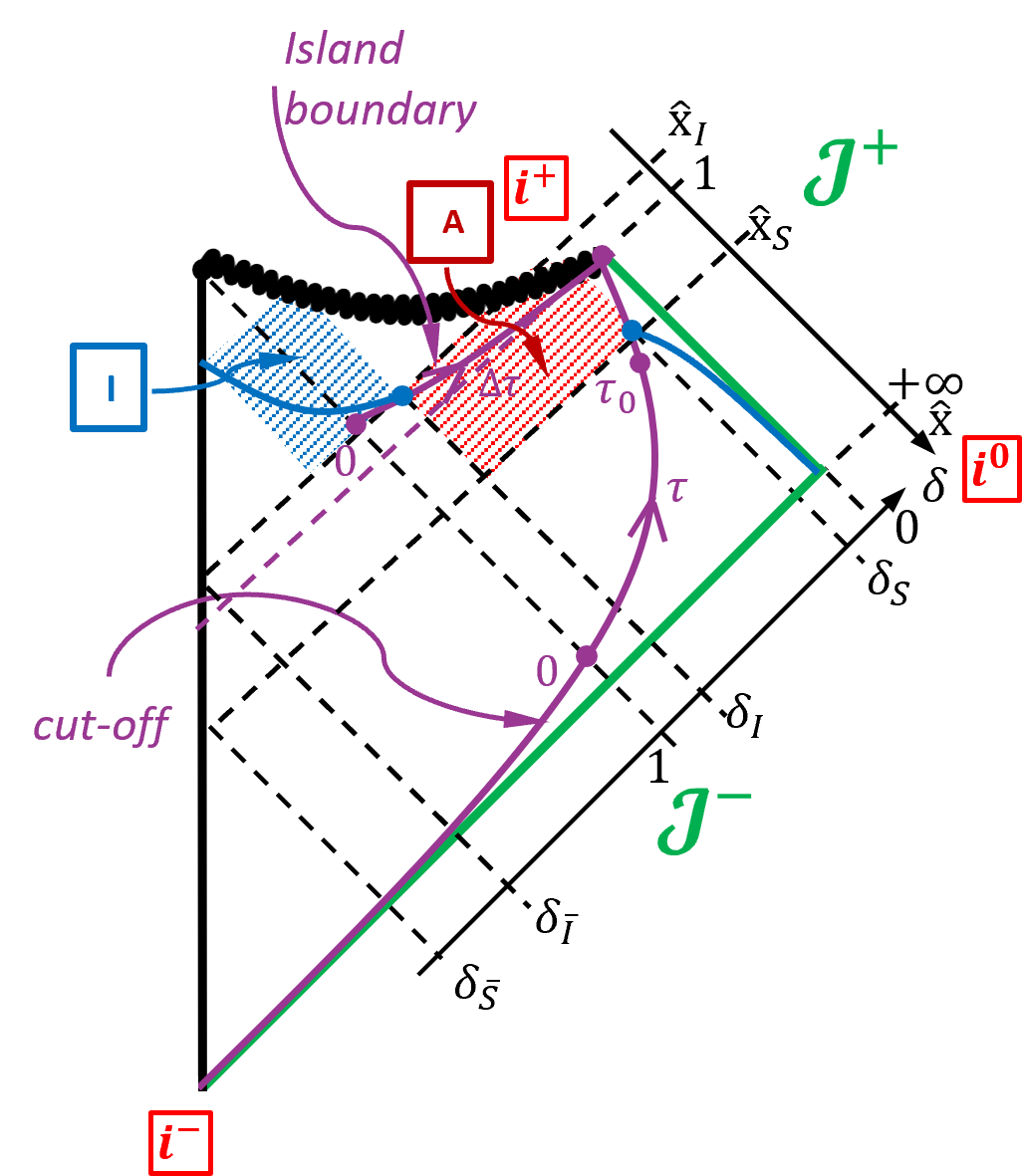}
    \caption{Position of the region ($A$) in which the inaccessible degrees of freedom live when an island ($I$) is present in the classical case is shown in red. The cut-off hypersurface is depicted in purple. There is also a $\tau$-axis along the cut-off surface. The island-boundary hypersurface is depicted in purple, as well. The $\Delta\tau$-axis are shown along this hypersurface. The degrees of freedom of the radiation live in a  causal diamond over a constant-time slice as well as in the island region. Both are shown in blue. Coordinates of all relevant points ($\delta_{S}$, $\delta_{I}$, $\hat{\mathrm{x}}_{S}$, $\hat{\mathrm{x}}_{I}$), as well as the ones obtained by the reflection from the boundary of space-time ($\delta_{\bar{S}}$, $\delta_{\bar{I}}$), are presented on both $\delta$-axis and $\hat{\mathrm{x}}$-axis. Note that the region III does not exist since the quantum-corrections have not been included.}\label{sl3}
    \end{center}
\end{figure}

In the presence of an island, the formula for the entropy of quantum fields is given by (\ref{ent_island}). It consists of two disjoint intervals along $\mathcal{J}^{-}$ hypersurface:
\begin{widetext}
    \begin{equation}
        S_{\mathrm{matter}}=\frac{1}{12}\ln{\frac{(\sigma^{+}_{S}-\sigma^{+}_{I})^{2}(\sigma^{+}_{\bar{S}}-\sigma^{+}_{\bar{I}})^{2}(\sigma^{+}_{\bar{S}}-\sigma^{+}_{S})^{2}(\sigma^{+}_{\bar{I}}-\sigma^{+}_{I})^{2}}{\epsilon^{4}(\sigma^{+}_{\bar{I}}-\sigma^{+}_{S})^{2}(\sigma^{+}_{\bar{S}}-\sigma^{+}_{I})^{2}e^{-2\rho(\sigma_{S})}e^{-2\rho(\sigma_{I})}}},\label{island}
    \end{equation}
\end{widetext}
where $\sigma_{\bar{I}}^{+}=\sigma_{I}^{-}$ and $\sigma_{\bar{S}}^{+}=\sigma_{S}^{-}$. To check if the formula (\ref{island}) holds when the island disappears, we take the limit $\sigma^{\pm}_{I}\to\sigma^{\pm}_{M}$. Since there exists a UV cut-off, lengths should not tend to zero, rather they should tend to that UV cut-off. Taking the limit $\sigma^{\pm}_{I}\to\sigma^{\pm}_{M}$, we have:
\begin{equation}
    \epsilon^{2}=\lim_{I\to M}g_{\mu\nu}(\sigma^{\mu}_{I}-\sigma^{\mu}_{M})(\sigma^{\nu}_{I}-\sigma^{\nu}_{M}).
\end{equation}
The island hypersurface is defined by: $\sigma^{+}+\sigma^{-}=const.$ Knowing that $\sigma_{M}^{+}=\sigma_{M}^{-}=\frac{1}{2}(\sigma^{+}_{I}+\sigma^{-}_{I})$, it is easy to show that (\ref{island}) reduces to (\ref{no_island}) when taking the limit $I\to M$. The same regularization should be used as in (\ref{S_FG_no_island_class}). The (\ref{island}) depends on $\tau$ and the position of the island. The dependence on $\tau$ manifests itself through $\sigma^{+}_{S}=\sigma^{+}_{0}+\tau$ and $\sigma^{-}_{S}=\sigma^{+}_{0}-2a\hat{\mathrm{x}}_{S}(\tau)$. This scenario is shown in Figure \ref{sl3}.
\par Next, we add the island area term. Since the DREH model has been derived by the method of dimensional reduction from the $4D$ Einstein-Hilbert action, the area term is given by a standard Bekenstein-Hawking entropy:
\begin{equation}
    \frac{A[\partial I]}{4G^{(4)}_{\text{N}}}=\frac{4\pi \lambda^{2}a^{2}}{4G\hbar}\mathrm{x}_{I}^{2}=\frac{(\lambda a)^{2}}{12\varepsilon}\mathrm{x}_{I}^{2}.\label{area_term}
\end{equation}
Now, the formula for the generalized entropy is given by the following expression:
\begin{widetext}
    \begin{equation}
        S_{gen}(\tau,\sigma_{I}^{+},\sigma^{-}_{I})=\frac{(\lambda a)^{2}}{12\varepsilon}\mathrm{x}_{I}^{2}+\frac{1}{12}\ln{\left[\frac{4a^{2}(b-a)\hat{\mathrm{x}}_{S}\left(\frac{b_{*}}{a}-\ln{(\hat{\mathrm{x}}_{S}-1)}\right)^{2}}{b^{3}(\hat{\mathrm{x}}_{S}-1)}\frac{(\sigma^{+}_{S}-\sigma^{+}_{I})^{2}(\sigma^{-}_{S}-\sigma^{-}_{I})^{2}(\sigma^{+}_{I}-\sigma^{-}_{I})^{2}}{\epsilon^{2}(\sigma^{+}_{S}-\sigma^{-}_{I})^{2}(\sigma^{+}_{I}-\sigma^{-}_{S})^{2}e^{-2\rho(\sigma_{I})}}\right]}.\label{S_gen_island_class}
    \end{equation}
\end{widetext}
The dependence of $\mathrm{x}_{I}(\sigma^{+}_{I},\sigma^{-}_{I})$ is set by equation (\ref{colaps_>_x_resenje}), while the dependence of $\rho(\sigma^{+}_{I},\sigma^{-}_{I})$ is defined in (\ref{kolaps_metrika_resenje}). The $\tau$ dependence comes from $\hat{\mathrm{x}}_{S}=\hat{\mathrm{x}}_{S}(\tau)$ (trough equation (\ref{cutoff_class})). According to the island formula (\ref{Island_formula}), the function (\ref{S_gen_island_class}) should be extramized with respect to the position of the island $(\sigma^{+}_{I},\sigma^{-}_{I})$. Using equations (\ref{colaps_>_x_resenje}) and (\ref{kolaps_metrika_resenje}), the extramization procedure yields the following equations:
\begin{widetext}
    \begin{align}
        1-\mathrm{x}_{I}&=\tilde{\varepsilon}\left[\frac{1}{2\mathrm{x}_{I}^{2}}+\frac{\hat{\mathrm{x}}_{S}-\ln{\delta}_{S}}{(\ln{\delta}_{I}-\ln{\delta_{S}})(\ln{\delta_{I}}-\hat{\mathrm{x}}_{S})}+\frac{1}{\hat{\mathrm{x}}_{I}-\ln{\delta_{I}}}\right],\label{ext_class_+}\\
        1-\mathrm{x}_{I}&=\tilde{\varepsilon}\left[\frac{1}{2\mathrm{x}_{I}^{2}}-\frac{1}{2\hat{\mathrm{x}}_{I}^{2}}+\frac{\hat{\mathrm{x}}_{I}-1}{\hat{\mathrm{x}}_{I}}\left(\frac{\hat{\mathrm{x}}_{S}-\ln{\delta}_{S}}{(\ln{\delta}_{I}-\ln{\delta_{S}})(\ln{\delta_{I}}-\hat{\mathrm{x}}_{S})}+\frac{1}{\hat{\mathrm{x}}_{I}-\ln{\delta_{I}}}\right)\right],\label{ext_class_-}
    \end{align}
\end{widetext}
where we have introduced $\tilde{\varepsilon}=\frac{\varepsilon}{(\lambda a)^{2}}$. First, we examine where the island appears along the $\delta_{I}=1$ hypersurface. After setting $\delta_{I}=1$ and $\hat{\mathrm{x}}_{I}=\mathrm{x}_{I}$, equations (\ref{ext_class_-}-\ref{ext_class_+}) reduce to:
\begin{align}
    1-\mathrm{x}_{I}&=\tilde{\varepsilon}\left[\frac{1}{2\mathrm{x}_{I}^{2}}+\frac{1}{\mathrm{x}_{I}}-\frac{\ln{\delta_{S}}-\hat{\mathrm{x}}_{S}}{\hat{\mathrm{x}}_{S}\ln{\delta_{S}}}\right],\label{ext_class_+_pocetak}\\
    \mathrm{x}_{I}^{2}&=\tilde{\varepsilon}\left[-1+\frac{\ln{\delta_{S}}}{\mathrm{x}_{I}-\ln{\delta_{S}}}-\frac{\hat{\mathrm{x}}_{S}}{\mathrm{x}_{I}-\hat{\mathrm{x}}_{S}}\right].\label{ext_class_-_pocetak}
\end{align}
Equation (\ref{ext_class_+_pocetak}) is a cubic equation for $\mathrm{x}_{I}$, and as such can be solved exactly. However, a perturbative solution up to the second order with respect to $\tilde{\varepsilon}$ should suffice. It is given by:
\begin{equation}
    \mathrm{x}_{I}=1-\tilde{\varepsilon}\left(\frac{3}{2}-\zeta\right)(1+2\tilde{\varepsilon}),\label{x_sol_mu}
\end{equation}
where we have defined $\zeta=\frac{\ln{\delta_{S}}-\hat{\mathrm{x}}_{S}}{\hat{\mathrm{x}}_{S}\ln{\delta_{S}}}$. Expressing $\delta_{S}=\delta_{S}(\hat{\mathrm{x}}_{S},\zeta)$ and substituting the solution (\ref{x_sol_mu}) into the second equation (\ref{ext_class_-_pocetak}) we arrive at the equation for $\hat{\mathrm{x}}_{S}$. The subsequent solution to this equation is given by:
\begin{equation}
    \hat{\mathrm{x}}_{S}=1+\tilde{\varepsilon}\left(\zeta-\frac{1}{2}\right)+\tilde{\varepsilon}^{2}\left(\zeta-\frac{3}{2}-\frac{1}{\zeta}\right).\label{x_hat_s_sol_mu}
\end{equation}
Finally, equation (\ref{cutoff_class}) becomes equation for $\zeta$:
\begin{equation}
    \zeta=\frac{1}{\hat{\mathrm{x}}_{S}}+\frac{1}{\frac{b_{*}}{a}-\hat{\mathrm{x}}_{S}-\ln{(\hat{\mathrm{x}}_{S}-1)}}.\label{mu1}
\end{equation}
Since $\frac{b_{*}}{a}\tilde{\varepsilon}\gg1$ the second term in (\ref{mu1}) can be neglected. Then we are left with the following:
\begin{equation}
    \zeta=1-\frac{\tilde{\varepsilon}}{2}.\label{mu2}
\end{equation}
Substituting the expression for $\zeta$ (\ref{mu2}) into equations (\ref{x_sol_mu}) and (\ref{x_hat_s_sol_mu}) and taking into account that $\frac{\tau_{0}}{2a}=-\ln{\delta_{S}}$, we arrive at the following result for the position of the island along the $\delta_{I}=1$ hypersurface:
\begin{align}
    \delta_{I}&=1,\label{delta_I_class_pocetak}\\
    \mathrm{x}_{I}&=1-\frac{\tilde{\varepsilon}}{2}(1+3\tilde{\varepsilon})=\hat{\mathrm{x}}_{I},\label{x_I_class_pocetak}\\
    \hat{\mathrm{x}}_{S}&=1+\frac{\tilde{\varepsilon}}{2}(1-4\tilde{\varepsilon}),\label{x_hat_S_class_pocetak}\\
    \frac{\tau_{0}}{2a}&=\frac{b_{*}-a}{a}-\ln{\frac{\tilde{\varepsilon}}{2}}+\frac{7}{2}\tilde{\varepsilon}=-\ln{\delta_{S}^{(0)}}.\label{tau_0}
\end{align}
Note that equation (\ref{tau_0}) implies that the first appearance of the island (along the $\delta_{I}=1$ hypersurface happens at late times, since $\tau_{0}/a\gg1/\tilde{\varepsilon}$). This fact implies that we are allowed to take the $|\ln{\delta_{S}}|\to\infty$ limit within equations (\ref{ext_class_+}) and (\ref{ext_class_-}). Those equations now become:
\begin{align}
    1-\mathrm{x}_{I}&=\tilde{\varepsilon}\left[\frac{1}{2\mathrm{x}_{I}^{2}}-\frac{1}{\hat{\mathrm{x}}_{S}-\ln{\delta_{I}}}+\frac{1}{\hat{\mathrm{x}}_{I}-\ln{\delta_{I}}}\right],\label{ext_class_+_late_time}\\
    1-\mathrm{x}_{I}&=\tilde{\varepsilon}\bigg{[}\frac{1}{2\mathrm{x}_{I}^{2}}-\frac{1}{2\hat{\mathrm{x}}_{I}^{2}}\nonumber\\
    &\hspace{10mm}+\frac{\hat{\mathrm{x}}_{I}-1}{\hat{\mathrm{x}}_{I}}\left(\frac{1}{\hat{\mathrm{x}}_{I}-\hat{\mathrm{x}}_{S}}+\frac{1}{\hat{\mathrm{x}}_{I}-\ln{\delta_{I}}}\right)\bigg{]}.\label{ext_class_-_late_time}
\end{align}
Guided by the expressions for the position of the island along the $\delta_{I}=1$ hypersurface (\ref{delta_I_class_pocetak}-\ref{tau_0}) we assume $\mathrm{x}_{I}=1-\xi$, $\hat{\mathrm{x}}_{I}=1-\hat{\xi}$ and $\hat{\mathrm{x}}_{S}=1+\mu$ for the position of the island at arbitrary time $\tau\geqslant\tau_{0}$. Here, $\xi$, $\hat{\xi}$, and $\mu$ are small functions of $\delta_{I}$, expanded to second order with respect to $\tilde{\varepsilon}$. Equations (\ref{ext_class_+_late_time}-\ref{ext_class_-_late_time}) now take the following form:
\begin{align}
    \xi&=\tilde{\varepsilon}\left[\frac{1}{2}+\xi+\frac{\mu+\hat{\xi}}{(1-\ln{\delta_{I}})^{2}}\right],\label{Ext1_class}\\
    \xi&=\tilde{\varepsilon}\left[\xi-\hat{\xi}+\frac{\hat{\xi}}{\hat{\xi}+\mu}(1+\hat{\xi})-\frac{\hat{\xi}}{1-\ln{\delta_{I}}}\right].\label{Ext2_class}
\end{align}
Equating (\ref{Ext1_class}) and (\ref{Ext2_class}) we get:
\begin{equation}
    \frac{1}{2}+\hat{\xi}+\frac{\mu+\hat{\xi}}{(1-\ln{\delta_{I}})^{2}}=\frac{\hat{\xi}}{\hat{\xi}+\mu}(1+\hat{\xi})-\frac{\hat{\xi}}{1-\ln{\delta_{I}}}.\label{ll1}
\end{equation}
The only term of the order $\mathcal{O}(1)$ on the right-hand side of equation (\ref{ll1}) is $\hat{\xi}/(\hat{\xi}+\mu)$; we conclude that it must be of the form:
\begin{equation}
    \frac{\hat{\xi}}{\hat{\xi}+\mu}=\frac{1}{2}+\eta,\label{ll2}
\end{equation}
where $\eta$ is a new variable of $\mathcal{\tilde{\varepsilon}}$ order. Placing the expression (\ref{ll2}) in equation (\ref{ll1}) yields the following:
\begin{equation}
    \eta=\frac{1}{2}\hat{\xi}+\frac{\hat{\xi}}{1-\ln{\delta_{I}}}+\frac{\mu+\hat{\xi}}{(1-\ln{\delta_{I}})^{2}}.\label{ll3}
\end{equation}
Using equations (\ref{ll2}) and (\ref{ll3}) it is easy to express $\xi$ and $\hat{\xi}$ in terms of $\mu$ and $\delta_{I}$:
\begin{align}
    \xi&=\frac{\tilde{\varepsilon}}{2}(1+\tilde{\varepsilon})\left[1+\frac{4\mu}{(1-\ln{\delta_{I}})^{2}}\right],\label{ll4}\\
    \hat{\xi}&=\mu\left[1+4\mu\left(\frac{1}{2}+\frac{1}{1-\ln{\delta_{I}}}+\frac{2}{(1-\ln{\delta_{I}})^{2}}\right)\right].\label{ll5}
\end{align}
The next step is to calculate $\delta_{I}=\delta_{I}(\mu)$ using equation (\ref{colaps_>_x_resenje}). Since there are logarithmic terms in this equation, $\delta_{I}$ can be determined only up to first order in $\tilde{\varepsilon}$:
\begin{equation}
    \delta_{I}=\frac{\hat{\xi}}{\xi}(1+\xi-\hat{\xi}).\label{ll6}
\end{equation}
Substituting expressions (\ref{ll4}) and (\ref{ll5}) into (\ref{ll6}) the following formula for $\delta_{I}$ is derived:
\begin{equation}
    \delta_{I}=\frac{\mu}{\tilde{\varepsilon}}\left(2-\tilde{\varepsilon}\right)\left[1+4\mu\left(\frac{1}{4}+\frac{1}{1-\ln{\delta_{I}}}+\frac{1}{(1-\ln{\delta_{I}})^{2}}\right)\right].\label{ll7}
\end{equation}
Finally, an expression for $\mu$ in terms of the time of evaporation $\tau$ is derived from (\ref{cutoff_class}):
\begin{equation}
    \mu=\frac{\tilde{\varepsilon}}{2}\frac{\delta_{S}(\tau)}{\delta_{S}(\tau_{0})}\left(1-\frac{7}{2}\tilde{\varepsilon}\right)\left(1-\frac{\tilde{\varepsilon}}{2}\frac{\delta_{S}(\tau)}{\delta_{S}(\tau_{0})}\right).\label{ll8}
\end{equation}
Using equations (\ref{ll4}-\ref{ll8}), the position of the island, in terms of the time since the formation of the island, $\Delta\tau=\tau-\tau_{0}$, is given by:
\begin{align}
    \delta_{I}&=e^{-\frac{\Delta\tau}{2a}}(1-4\tilde{\varepsilon})\left[1+2\tilde{\varepsilon}\frac{2+\frac{\Delta\tau}{2a}}{\left(1+\frac{\Delta\tau}{2a}\right)^{2}}e^{-\frac{\Delta\tau}{2a}}\right]\label{delta_I_class},\\
    \hat{\mathrm{x}}_{I}&=1-\frac{\tilde{\varepsilon}}{2}e^{-\frac{\Delta\tau}{2a}}\left(1-\frac{7}{2}\tilde{\varepsilon}\right)\nonumber\\
    &\hspace{10mm}\times\left[1+\frac{\tilde{\varepsilon}}{2}\left(1+4\frac{3+\frac{\Delta\tau}{2a}}{\left(1+\frac{\Delta\tau}{2a}\right)^{2}}\right)e^{-\frac{\Delta\tau}{2a}}\right],\label{x_hat_I_class}\\
    \mathrm{x}_{I}&=1-\frac{\tilde{\varepsilon}}{2}(1+\tilde{\varepsilon})\left[1+\frac{2\tilde{\varepsilon}}{\left(1+\frac{\Delta\tau}{2a}\right)^{2}}e^{-\frac{\Delta\tau}{2a}}\right],\label{x_I_class}\\
    \hat{\mathrm{x}}_{S}&=1+\frac{\tilde{\varepsilon}}{2}e^{-\frac{\Delta\tau}{2a}}\left(1-\frac{7}{2}\tilde{\varepsilon}\right)\left[1-\frac{\tilde{\varepsilon}}{2}e^{-\frac{\Delta\tau}{2a}}\right]\label{x_hat_S_class}.
\end{align}
We can conclude that the island appears behind the horizon, which is expected. Notice that taking the limit $\Delta\tau\to0$ of equations (\ref{delta_I_class}-\ref{x_hat_S_class}) reduces them to equations (\ref{delta_I_class_pocetak}-\ref{x_hat_S_class_pocetak}). In addition, taking the limit $\Delta\tau\to\infty$ amounts to $\delta_{I}=\to0$, $\hat{\mathrm{x}}_{I}\to1$, which means that the island ends at future time infinity. This is expected, since there is no quantum correction. Also, $\hat{\mathrm{x}}\to1$. It is interesting to note that the island and the cut-off surface almost mirror the horizon at $\hat{\mathrm{x}}_{H}=1$.
\par Now we calculate the generalized entropy of the radiation (\ref{S_gen_island_class}) along the island hypersurface. The direct calculation yields the following result:
\begin{widetext} 
    \begin{equation}
        S_{RW}=\frac{(\lambda a)^{2}}{12\varepsilon}-\frac{1}{12}-\frac{\varepsilon}{24(\lambda a)^{2}}+\frac{1}{12}\ln{\left(\frac{\varepsilon}{2(\lambda a)^{2}}\frac{b-a}{b}\right)}+\frac{1}{6}\ln{\frac{-8a^{2}\ln{\delta_{S}^{(0)}}}{\epsilon^{2}}}-\frac{\varepsilon}{6(\lambda a)^{2}}\frac{1-\frac{\Delta\tau}{2a}}{\left(1+\frac{\Delta\tau}{2a}\right)^{2}}e^{-\frac{\Delta\tau}{2a}}.\label{S_FG_I}
    \end{equation}
\end{widetext}
Equation (\ref{S_FG_I}) contains a constant term as well as an exponentially decreasing term. The constant term should be regularized since it depends on the UV cut-off $\epsilon$. Neglecting the UV term in the constant part of (\ref{S_FG_I}), the dominant part is the Bekenstein-Hawking entropy $S_{BH}^{(0)}=\frac{(\lambda a)^{2}}{12\varepsilon}$. We conclude that the constant part can be regarded as quantum-corrected Bekenstein-Hawking entropy $S_{BH}^{\mathrm{corr}}$. Then,  fine-grained entropy at late times is given by the following formula:
\begin{widetext}
    \begin{equation}
        \boxed{S_{FG}=\min\left\{\frac{\Delta\tau}{24a}+\frac{1}{6}\ln{\frac{\tau}{2b}}-\frac{1}{12}\ln{\frac{\varepsilon}{2(\lambda a)^{2}}},S_{BH}^{\mathrm{corr}}-\frac{\varepsilon}{6(\lambda a)^{2}}\frac{1-\frac{\Delta\tau}{2a}}{\left(1+\frac{\Delta\tau}{2a}\right)^{2}}e^{-\frac{\Delta\tau}{2a}}\right\}},\label{S_{FG_class}}
    \end{equation}
\end{widetext}
From (\ref{S_{FG_class}}) we conclude that at late times the fine-grained entropy saturates at a value close to the Bekenstein-Hawking entropy of the black hole. This means that, without the back-reaction of quantum fields on the space-time geometry, we cannot reproduce the correct Page curve, and get the same result as in the case of an eternal black hole \cite{DREH1}.
\begin{figure}
    \begin{center}
    \includegraphics[width=8cm, height=4cm]{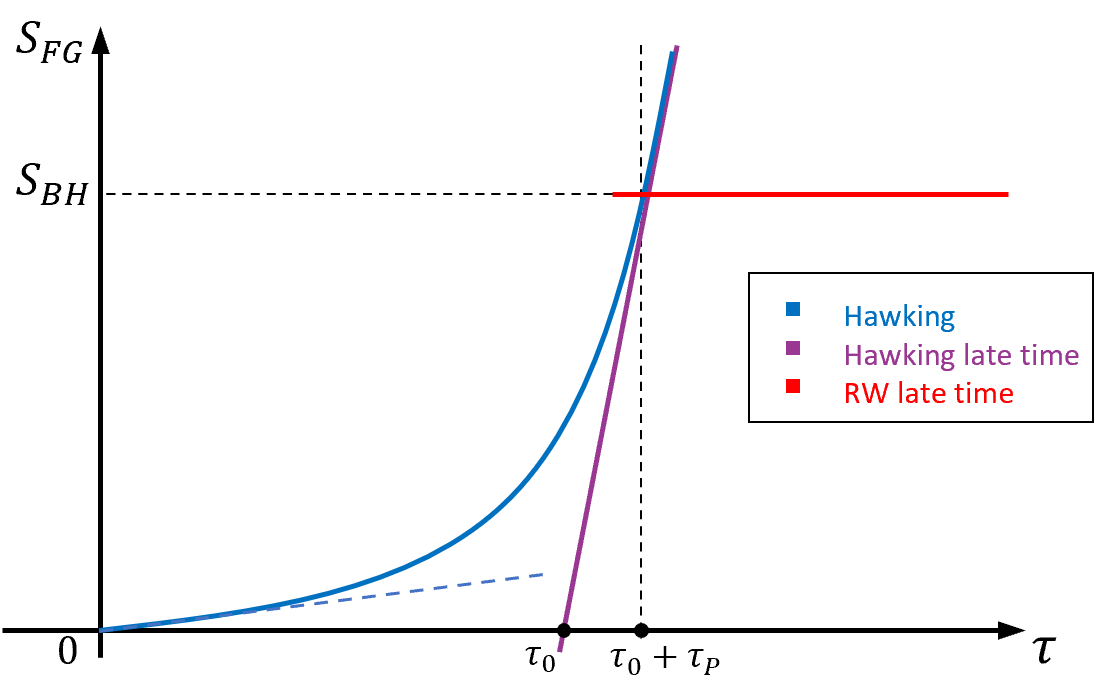}
    \caption{The Page curve for the classical collapse scenario is depicted here. The blue line represents the fine-grained entropy along the Hawking's saddle point, while the red line show the entropy along the replica-wormholes saddle point at late times. The purple line show the asymptotic behavior of the fine-grained entropy along the Hawking's saddle point at late times.}\label{sl4}
    \end{center}
\end{figure}
\par To find the time at which the phase transition occurs, i.e. the Page time, we equate the dominant parts within (\ref{S_{FG_class}}). The Page time is then given by:
\begin{equation}
    \tau_{P}=24aS_{BH}^{\text{corr}}.\label{t_page_class}
\end{equation}
The result (\ref{t_page_class}) coincides with the result obtained in the case of an eternal black hole \cite{DREH1}, since the connection between the time in \cite{DREH1} and the equation (\ref{t_page_class}) is $t_{P}=\frac{1}{2}\tau_{P}$. In Figure \ref{sl4} the Page curve for a classical collapse scenario is depicted.
\section{Page curve for the quantum-corrected collapse scenario}
Now we turn to the investigation of the Page curve in the case of the quantum-corrected collapse scenario. In the previous section, we have seen that the replica-wormholes saddle point does not generate the decreasing contribution to the fine-grained entropy needed for the Page curve in the case of an classical evaporating black hole. This section will also consist of two parts. First, we investigate the no-island case, and then proceed on to the island case.
\subsection{No-island case}
\begin{figure}
    \begin{center}
    \includegraphics[width=5.5cm, height=7cm]{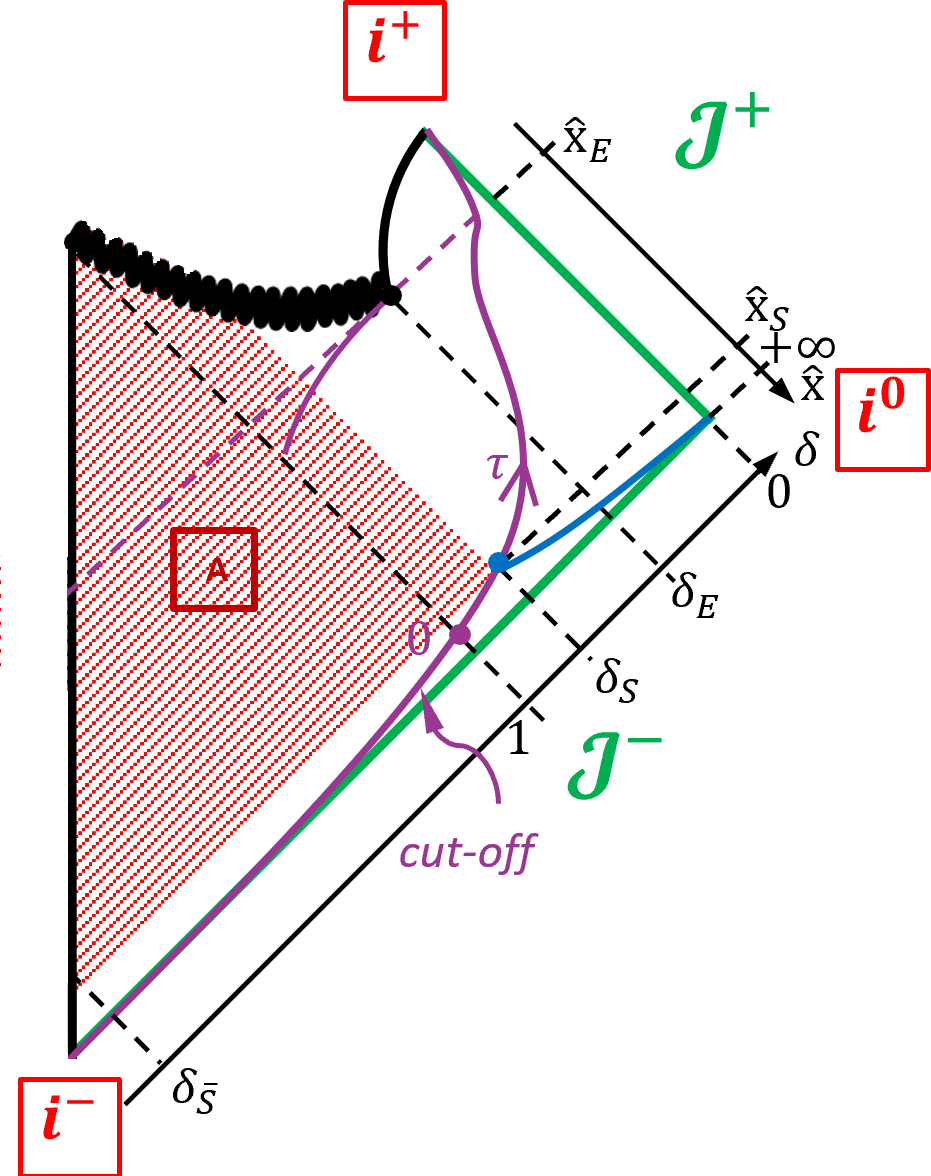}
    \caption{Position of the region ($A$) in which the inaccessible degrees of freedom live for a no-island case when the quantum-corrections are included is shown in red. The cut-off hypersurface is depicted in purple. There is also a $\tau$-axis along the cut-off surface. The degrees of freedom of the radiation live in a  causal diamond over a constant-time slice shown in blue. Coordinates of all relevant points ($\delta_{S}$, $\hat{\mathrm{x}}_{S}$), as well as the ones obtained by the reflection from the boundary of space-time ($\delta_{\bar{S}}$), are presented on both $\delta$-axis and $\hat{\mathrm{x}}$-axis.}\label{PD_q_no_island}
    \end{center}
\end{figure}

Here we analyze the impact of the back-reaction of quantum fields onto the fine-grained entropy calculated along the Hawking's saddle point. The correct Penrose diagram for this case is shown in Figure \ref{PD_q_no_island}. One does not expect drastic changes to the results obtained in section \ref{Sec_no_island_class}. The fine-grained entropy is given by the same formula (\ref{ent2}) as in the classical case:
\begin{equation}
    S_{\mathrm{matter}}=\frac{1}{12}\ln{\frac{(\sigma_{S}^{+}-\sigma_{\bar{S}}^{+})^{2}}{4\epsilon^{2}e^{-2\rho_{S}(\sigma)}}}.\label{no_island_q}
\end{equation}
In the same way as in section \ref{Sec_no_island_class} the cut-off surface (\ref{cutoff_class}) is defined by $r_{*}=b_{*}$. Using the formula (\ref{y_<_resenje}) for the tortoise coordinate, one derives the following expression:
\begin{equation}
    \frac{b_{*}}{a}-\frac{\tau}{2a}=\mathcal{J}(\hat{\mathrm{x}}_{S})-\frac{9}{2}-\frac{\tilde{\varepsilon}}{8}\left[(2\hat{\mathrm{x}}_{S}+3)\ln{\hat{\mathrm{x}}_{S}}+5\hat{\mathrm{x}}_{S}-2\mathrm{L}_{S}\right],\label{cutoff_q}
\end{equation}
where we have once again introduced the time of evaporation $-\ln{\delta_{S}}=\frac{\tau}{2a}$ in the same way as in (\ref{veza_tau_x_hat_S}). Note that we still assume $b_{*}\tilde{\varepsilon}\gg1$. Now we calculate the fine-grained entropy along the Hawking's saddle point, keeping in mind the regularization $S_{H}(0)=0$:
\begin{equation}
    S_{H}(\tau)=\frac{1}{12}\ln{\left(\frac{a^{2}}{b^{2}}\left(\frac{\tau}{2a}+\hat{\mathrm{x}}_{S}\right)^{2}\frac{\mathrm{F}^{+}_{S}\mathrm{F}^{-}_{S}e^{2\rho_{S}}}{\mathrm{F}^{-}_{S}\mathrm{F}^{+}_{S}}\right)}.\label{S_H_q}
\end{equation}
Let us examine the behavior of the fine-grained entropy (\ref{S_H_q}) at the beginning of the evaporation, i.e. $\tau\to0$, the same way as in the classical case \ref{Sec_no_island_class}. First, one has to solve the transcendental equation (\ref{cutoff_q}) for $\hat{\mathrm{x}}_{S}$ in terms of $\tau$. When $\tau=0$ we have: $\hat{\mathrm{x}}_{S}=\mathrm{x}_{S}=\frac{b}{a}$, which implies that for the small $\tau$, $\hat{\mathrm{x}}_{S}$ takes the following form: $\hat{\mathrm{x}}_{S}=\frac{b}{a}+\eta$, where $\eta$ is small. Expanding equation (\ref{cutoff_class}), we get the following solution:
\begin{equation}
    \hat{\mathrm{x}}_{S}^{(0)}=\frac{b}{a}-\frac{b-a}{2ab}\tau\left\{1-\frac{\tilde{\varepsilon}}{8}\left[2\ln{\left(1-\frac{a}{b}\right)}+\frac{2a}{b-a}-\frac{a^{2}}{b^{2}}\right]\right\}.\label{b1}
\end{equation}
After leaving only the dominant term in the $\frac{a}{b}\to0$ expansion, we are left with:
\begin{equation}
    \hat{\mathrm{x}}_{S}^{(0)}=\frac{b}{a}-\frac{b-a}{2ab}\tau\left(1-\frac{\tilde{\varepsilon}}{6}\frac{a^{3}}{b^{3}}\right).
\end{equation}
Compared with equation (\ref{a1}) we see that there is a small modification due to quantum corrections. The direct calculation of the fine-grained entropy (\ref{S_H_q}) yields:
\begin{equation}
    S_{H}^{(0)}(\tau)=\frac{a\tau}{8b^{2}}\left(1+\frac{\tilde{\varepsilon}}{4}\frac{b^{2}}{a^{2}}\right),\label{S_FG_early_times_q}
\end{equation}
where we have left only the dominant term in the $\frac{a}{b}\to0$ expansion. Compared with (\ref{S_FG_early_times_class}) once again, we see that there is a small quantum correction.
\par Next we move onto the late-time investigation of the formula (\ref{S_H_q}). Since at late times $\hat{\mathrm{x}}_{S}=1+\tilde{\varepsilon}\mu$, we need to expand equation (\ref{cutoff_q}) around $\hat{\mathrm{x}}_{S}=1$:
\begin{equation}
    \frac{b_{*}-a}{b_{*}}-\frac{\tau}{2a}=\left(1+\frac{11}{8}\tilde{\varepsilon}\right)\frac{4}{\tilde{\varepsilon}}\left(e^{\frac{\tilde{\varepsilon}}{4}\ln{\left(\tilde{\varepsilon}\left(\mu-\frac{1}{4}\right)\right)}}-1\right)+\tilde{\varepsilon}\mu.\label{x_hat_s_late_times}
\end{equation}
The solution to equation (\ref{x_hat_s_late_times}) is given by the following expression:
\begin{equation}
    \hat{\mathrm{x}}_{S}^{(\infty)}=1+\frac{\tilde{\varepsilon}}{4}\left(1+3e^{-\frac{\tau-\tau_{0}}{2\bar{a}}}\right),\label{b2}
\end{equation}
where $\tau_{0}$ is defined later via equation (\ref{tau_0_q}). We will use the notation: $\Delta\tau=\tau-\tau_{0}$. Since, at late times, the metric is asymptotically flat, we can write $\mathrm{F}^{+}_{S}\mathrm{F}^{-}_{S}e^{2\rho_{S}}\approx-1$. When $\hat{\mathrm{x}}_{S}\approx1$, the coordinate transformations become:
\begin{equation}
    \mathrm{F}^{-}_{S}\approx\hat{\mathrm{x}}^{(\infty)}_{S}-1-\frac{\tilde{\varepsilon}}{4}=\frac{3}{4}\tilde{\varepsilon}e^{-\frac{\Delta\tau}{2\bar{a}}}.\label{b3}
\end{equation}
Inserting (\ref{b2}) and (\ref{b3}) into equation (\ref{S_H_q}) yields:
\begin{equation}
    S_{H}^{(\infty)}(\tau)=\frac{\tau-\tau_{0}}{24\bar{a}}+\frac{1}{6}\ln{\frac{\tau}{2b}}-\frac{1}{12}\ln{\left(-\frac{3\tilde{\varepsilon}}{4}\mathrm{F}^{+}_{\infty}\right)},\label{S_FG_late_times_q}
\end{equation}
where $\mathrm{F}^{+}_{\infty}$ is the value of the coordinate transformation at late times $\sigma^{+}\to\infty$. We can conclude that in the case of the Hawking's saddle point we have linear growth of the fine-grained entropy at late times. That is an expected result. Compared with (\ref{S_FG_late_times_class}) we can see that the entropy behaves in the same way as in the classical case; with quantum corrected surface gravity later given by equation (\ref{kappa}).
\subsection{Island case}
\begin{figure}
    \begin{center}
    \includegraphics[width=6.5cm, height=7cm]{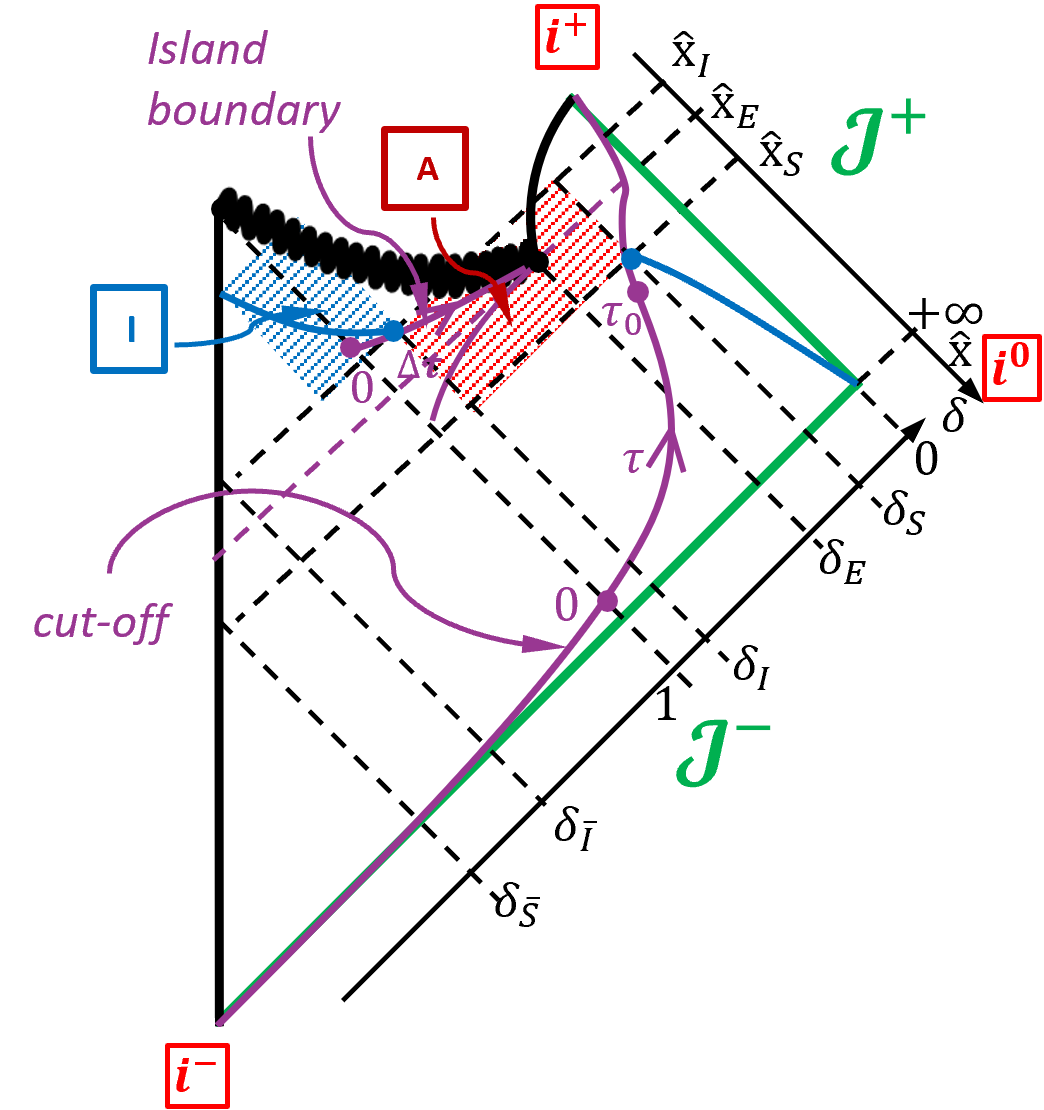}
    \caption{Position of the region ($A$) in which the inaccessible degrees of freedom live when an island ($I$) is present in the quantum-corrected case is shown in red. The cut-off hypersurface is depicted in purple. There is also a $\tau$-axis along the cut-off surface. The island-boundary hypersurface is depicted in purple, as well. The $\Delta\tau$-axis is shown along this hypersurface. The degrees of freedom of the radiation live in a  causal diamond over a constant-time slice as well as in the island region. Both are shown in blue. Coordinates of all relevant points ($\delta_{S}$, $\delta_{I}$, $\hat{\mathrm{x}}_{S}$, $\hat{\mathrm{x}}_{I}$, $\hat{\mathrm{x}}_{E}$, $\delta_{E}$), as well as the ones obtained by the reflection from the boundary of space-time ($\delta_{\bar{S}}$, $\delta_{\bar{I}}$), are presented on both $\delta$-axis and $\hat{\mathrm{x}}$-axis.}\label{PD_q_island}
    \end{center}
\end{figure}

Now we turn to the investigation of the replica-wormholes saddle point. The same formulas (\ref{area_term}) and (\ref{island}) hold for the parts of the generalized entropy in the case of the quantum-corrected geometry. The Penrose diagram in this case is shown in Figure \ref{PD_q_island}. Extremization yields the following equations for the position of the island: 
\begin{widetext}  
    \begin{align}
        \mathrm{x}_{I}\partial_{+}\mathrm{x}_{I}+\tilde{\varepsilon}\partial_{+}\rho_{I}&=-\frac{\tilde{\varepsilon}}{2a}\left[\frac{\hat{\mathrm{x}}_{S}-\ln{\delta}_{S}}{(\ln{\delta}_{I}-\ln{\delta_{S}})(\ln{\delta_{I}}-\hat{\mathrm{x}}_{S})}+\frac{1}{\hat{\mathrm{x}}_{I}-\ln{\delta_{I}}}\right],\label{+_j_na}\\
        \mathrm{x}_{I}\partial_{-}\mathrm{x}_{I}+\tilde{\varepsilon}\partial_{-}\rho_{I}&=\frac{\tilde{\varepsilon}}{2a}\left[\frac{\hat{\mathrm{x}}_{S}-\ln{\delta}_{S}}{(\hat{\mathrm{x}}_{I}-\ln{\delta_{S}})(\hat{\mathrm{x}}_{I}-\hat{\mathrm{x}}_{S})}+\frac{1}{\hat{\mathrm{x}}_{I}-\ln{\delta_{I}}}\right].\label{-_j_na}
    \end{align}
\end{widetext}
At the beginning, we examine the appearance of the island along the $\delta_{I}=1$ hypersurface, in the same way as we did in section \ref{Sec_island_class}. Note that along this hypersurface we have a continuity condition (see \cite{DREH2}), which implies:
\begin{align}
    \mathrm{x}_{I}\partial_{+}\mathrm{x}_{I}+\tilde{\varepsilon}\partial_{+}\rho_{I}&=\frac{\mathrm{x}_{I}-1}{2a}\\
    \mathrm{x}_{I}\partial_{-}\mathrm{x}_{I}+\tilde{\varepsilon}\partial_{-}\rho_{I}&=-\frac{\mathrm{x}_{I}}{2a}
\end{align}
Applying these continuity conditions, equations (\ref{+_j_na}) and (\ref{-_j_na}) become:
\begin{align}
    1-\mathrm{x}_{I}&=\tilde{\varepsilon}\left[\frac{1}{\mathrm{x}_{I}}-\frac{\ln{\delta_{S}}-\hat{\mathrm{x}}_{S}}{\hat{\mathrm{x}}_{S}\ln{\delta_{S}}}\right],\label{ext_qm_+_pocetak}\\
    \mathrm{x}_{I}^{2}&=\tilde{\varepsilon}\left[-1+\frac{\ln{\delta_{S}}}{\mathrm{x}_{I}-\ln{\delta_{S}}}-\frac{\hat{\mathrm{x}}_{S}}{\mathrm{x}_{I}-\hat{\mathrm{x}}_{S}}\right].\label{ext_qm_-_pocetak}
\end{align}
Notice that the second equation (\ref{ext_qm_-_pocetak}) is exactly the same as in the classical case (\ref{ext_class_-_pocetak}); while the other equation (\ref{ext_qm_+_pocetak}) is simpler than the corresponding classical equation (\ref{ext_class_+_pocetak}); now it is a simple quadratic equation. Once again, we define $\zeta=\frac{\ln{\delta_{S}}-\hat{\mathrm{x}}_{S}}{\hat{\mathrm{x}}_{S}\ln{\delta_{S}}}$. Both equations can be exactly solved in terms of $\zeta$. The solution is given by:
\begin{align}
    \mathrm{x}_{I}&=\frac{1+\tilde{\varepsilon}\zeta}{2}\left[1+\sqrt{1-\frac{4\tilde{\varepsilon}}{(1+\tilde{\varepsilon}\zeta)^{2}}}\right],\label{qq1}\\
    \hat{\mathrm{x}}_{S}&=\mathrm{x}_{I}\frac{1+\frac{1}{2}\zeta\mathrm{x}_{I}\left(1+\sqrt{1+\frac{4\tilde{\varepsilon}}{\zeta\mathrm{x}^{2}_{I}(1+\tilde{\varepsilon}\zeta)}}\right)}{1+\frac{\zeta\mathrm{x}_{I}^{2}}{1+\tilde{\varepsilon}\zeta}}.\label{qq2}
\end{align}
To get some quantitative results, let us expand equations (\ref{qq1}) and (\ref{qq2}) in terms of $\tilde{\varepsilon}$ up to the second order:
\begin{align}
    \mathrm{x}_{I}&=1+\tilde{\varepsilon}(1+\tilde{\varepsilon})(\zeta-1),\label{qq11}\\
    \hat{\mathrm{x}}_{S}&=1+\tilde{\varepsilon}\left(\zeta-\frac{\tilde{\varepsilon}}{\zeta}\right).\label{qq22}
\end{align}
Now, we calculate $\zeta=\frac{1}{\hat{\mathrm{x}}_{S}}-\frac{1}{\ln{\delta_{S}}}$. As in the classical case, $\ln{\delta_{S}}\gg1$ and we can neglect the second term. Then we have $\zeta=1-\tilde{\varepsilon}(1+\tilde{\varepsilon})$. Finally, one has to find the expression for $\tau_{0}$ using equation (\ref{cutoff_q}). Substituting $\zeta$ into equations (\ref{qq11}) and (\ref{qq22}), the position of the island along the $\delta_{I}=1$ hypersurface is given by:
\begin{align}
    \delta_{I}&=1,\label{delta_I_q_pocetak}\\
    \mathrm{x}_{I}&=1-\tilde{\varepsilon}^{2}=\hat{\mathrm{x}}_{I},\label{x_I_q_pocetak}\\
    \hat{\mathrm{x}}_{S}&=1+\tilde{\varepsilon}(1-2\tilde{\varepsilon}),\label{x_hat_S_q_pocetak}\\
    \frac{\tau_{0}}{2a}&=\frac{b_{*}-a}{a}-\left(1+\frac{11}{8}\tilde{\varepsilon}\right)\frac{4}{\tilde{\varepsilon}}\left(e^{\frac{\tilde{\varepsilon}}{4}\ln{\frac{3\tilde{\varepsilon}}{4}}}-1\right)-\tilde{\varepsilon}.\label{tau_0_q}
\end{align}
Comparing equations (\ref{delta_I_q_pocetak}-\ref{tau_0_q}) to the classical result (\ref{delta_I_class_pocetak}-\ref{tau_0}) we see that the first correction for the position of the island $\mathrm{x}_{I}$ along the $\delta_{I}=1$ hypersurface is of $\mathcal{O}(\tilde{\varepsilon}^{2})$ order opposite to the $\mathcal{O}(\tilde{\varepsilon})$ order in the classical case. Note that the time when the island forms $\tau_{0}$ is the same as in the classical result up to the $\mathcal{O}(1)$ order. The change in higher orders is exacted since we have added the quantum corrections to the metric.
\par Now we turn to finding the position of the island at arbitrary time $\tau>\tau_{0}$. Since $b_{*}\tilde{\varepsilon}\gg1$, we can neglect terms of type $1/(\ln{\delta_{S}})$, as we did in a classical case. Equations (\ref{+_j_na}) and (\ref{-_j_na}) now take the following form:
\begin{align}
    \mathrm{x}_{I}\partial_{+}\mathrm{x}_{I}+\tilde{\varepsilon}\partial_{+}\rho_{I}&=\frac{\tilde{\varepsilon}}{2a}\left[\frac{1}{\hat{\mathrm{x}}_{S}-\ln{\delta_{I}}}-\frac{1}{\hat{\mathrm{x}}_{I}-\ln{\delta_{I}}}\right],\label{+_j_na_late_time}\\
    \mathrm{x}_{I}\partial_{-}\mathrm{x}_{I}+\tilde{\varepsilon}\partial_{-}\rho_{I}&=\frac{\tilde{\varepsilon}}{2a}\left[\frac{1}{\hat{\mathrm{x}}_{I}-\hat{\mathrm{x}}_{S}}+\frac{1}{\hat{\mathrm{x}}_{I}-\ln{\delta_{I}}}\right].\label{-_j_na_late_time}
\end{align}
The expected form of the solution is: $\mathrm{y}_{I}=1+\xi$, $\hat{\mathrm{x}}_{I}=1+\hat{\xi}$, and $\hat{\mathrm{x}}_{S}=1+\mu$, where $\xi$, $\hat{\xi}$ and $\mu$ are of $\mathcal{O}(\tilde{\varepsilon})$ order. Substituting this ansatz into the expressions (\ref{izvod_x_+}-\ref{izvod_rho_-}) for the derivatives we get:
\begin{align}
    \partial_{+}\mathrm{x}_{I}&=\frac{1}{2a}\left(1-\frac{1}{\sqrt{\mathrm{y}_{I}^{2}-\tilde{\varepsilon}}}\right),\label{izv_x_+}\\
    \partial_{-}\mathrm{x}_{I}&=-\frac{1}{2a\mathrm{F}^{-}_I{}}\left(1-\frac{1-\frac{\tilde{\varepsilon}}{4}}{\sqrt{\mathrm{y}_{I}^{2}-\tilde{\varepsilon}}}\right),\label{izv_x_-}\\
    \partial_{+}\rho&=\frac{1}{4a\tilde{a}_{I}}\frac{\mathrm{y}_{I}}{(\mathrm{y}_{I}^{2}-\tilde{\varepsilon})^{\frac{3}{2}}},\label{izv_rho_+}\\
    \partial_{-}\rho&=-\frac{1}{4a\mathrm{F}^{-}_{I}}\left(\frac{\hat{\mathrm{x}}_{I}^{2}-1}{\hat{\mathrm{x}}_{I}^{2}}-\frac{1}{\tilde{a}_{I}}\frac{\mathrm{y}^{2}_{I}-1}{\mathrm{y}_{I}^{2}}\right),\label{izv_rho_-}
\end{align}
where $\mathrm{y}$ and $\tilde{a}$ are defined through equation (\ref{y_deff}). Using equations (\ref{izv_x_+}) and (\ref{izvod_rho_+}), expression (\ref{+_j_na_late_time}) reduces to $\mathrm{y}_{I}=1$, or $\xi=0$. Notice that we can only find the position of the island up to the $\mathcal{O}(\tilde{\varepsilon})$ order, since the solution is of the same order. This implies that the left-hand side of equation (\ref{+_j_na_late_time}) is equal to zero. The $\hat{\xi}$ is easily obtained using equation (\ref{resenje_y=1}). The result is:
\begin{equation}
    \hat{\xi}=\frac{1}{4}(1-\delta_{I}).\label{xi_hat_q}
\end{equation}
From equation (\ref{F_-_od_F_x_hat}) we conclude that:
\begin{equation}
    \mathrm{F}^{-}(\hat{\mathrm{x}}_{I})=\hat{\xi}-\frac{\tilde{\varepsilon}}{4}.\label{F_-_I}
\end{equation}
Placing expression (\ref{F_-_I}) in equation (\ref{-_j_na_late_time}) we get:
\begin{equation}
    \frac{\tilde{\varepsilon}}{4}=\tilde{\varepsilon}\frac{\hat{\xi}-\frac{\tilde{\varepsilon}}{4}}{\hat{\xi}-\mu}\label{j_na_-_sred}
\end{equation}
After solving equation (\ref{j_na_-_sred}) and taking into account that (\ref{xi_hat_q}), $\mu$ is given by:
\begin{equation}
    \mu=\frac{1}{4}\left(1+3\delta_{I}\right).\label{mu_q}
\end{equation}
Setting $\Delta\tau=\tau-\tau_{0}$ into equation (\ref{x_hat_s_late_times}) we arrive at the following equation for $\mu$:
\begin{equation}
    -\frac{\Delta\tau}{2a}\left(1-\frac{11}{8}\tilde{\varepsilon}\right)e^{-\frac{\tilde{\varepsilon}}{4}\ln{\frac{3\tilde{
    \varepsilon
    }}{4}}}=\frac{4}{\tilde{\varepsilon}}\left(e^{\frac{\tilde{\varepsilon}}{4}\ln{\frac{4\mu-1}{3}}}-1\right)+\tilde{\varepsilon}(\mu-1).\label{w1}
\end{equation}
Since there are logarithms on the right-hand side of equation (\ref{w1}), we can only be certain that it is correct up to the $\mathcal{O}(1)$ order. After taking the limit $\tilde{\varepsilon}\to0$ and solving the equation for $\mu$ we arrive at:
\begin{equation}
    \mu=\frac{1}{4}\left(1+3e^{-\frac{\Delta\tau}{2\bar{a}}}\right),\label{w2}
\end{equation}
where we have introduced a new constant $\bar{a}$, which can be interpreted trough the quantum-corrected surface gravity:
\begin{equation}
    \kappa=\frac{1}{2\bar{a}}=\frac{1}{2a}\left(1-\frac{11}{8}\tilde{\varepsilon}\right)e^{-\frac{\tilde{\varepsilon}}{4}\ln{\frac{3\tilde{\varepsilon}}{4}}}.\label{kappa}
\end{equation}
Combining equations (\ref{xi_hat_q}), (\ref{mu_q}) and (\ref{w2}) we arrive at the following expressions for the position of the island:
\begin{align}
    \delta_{I}&=e^{-\frac{\Delta\tau}{2\bar{a}}},\label{delta_I_q}\\
    \hat{\mathrm{x}}_{I}&=1+\frac{\tilde{\varepsilon}}{4}\left(1-e^{-\frac{\Delta\tau}{2\bar{a}}}\right),\label{x_hat_I_q}\\
    \mathrm{x}_{I}&=1-\frac{\tilde{\varepsilon}}{4}\frac{\Delta\tau}{2\bar{a}},\label{x_I_q}\\
    \hat{\mathrm{x}}_{S}&=1+\frac{\tilde{\varepsilon}}{4}\left(1+3e^{-\frac{\Delta\tau}{2\bar{a}}}\right).\label{x_hat_S_q}
\end{align}
Note that the island appears behind the event horizon as is expected in case of the evaporating black hole. Comparing equations (\ref{x_hat_AH_early}) and (\ref{x_hat_I_q}), we conclude that the island and the apparent horizon mirror each other with respect to the event horizon given by (\ref{x_H}). In addition, the island crosses the singularity a small portion of time before the end-point of the evaporation, as is the case of the other models of dilaton gravity \cite{BPPislands,RSTislands}. After taking the limit $\Delta\tau\to0$ equations (\ref{delta_I_q}-\ref{x_hat_S_q}) reduce to equations (\ref{x_I_q_pocetak}-\ref{tau_0_q}). In contrast to the classical case where $\mathrm{x}_{I}$ tends to the constant at late times (\ref{x_I_class}), here $\mathrm{x}_{I}$ decreases as time passes, which is the desired result from the perspective of reproducing the Page curve.
\par
\begin{figure}
    \begin{center}
    \includegraphics[width=8cm, height=4cm]{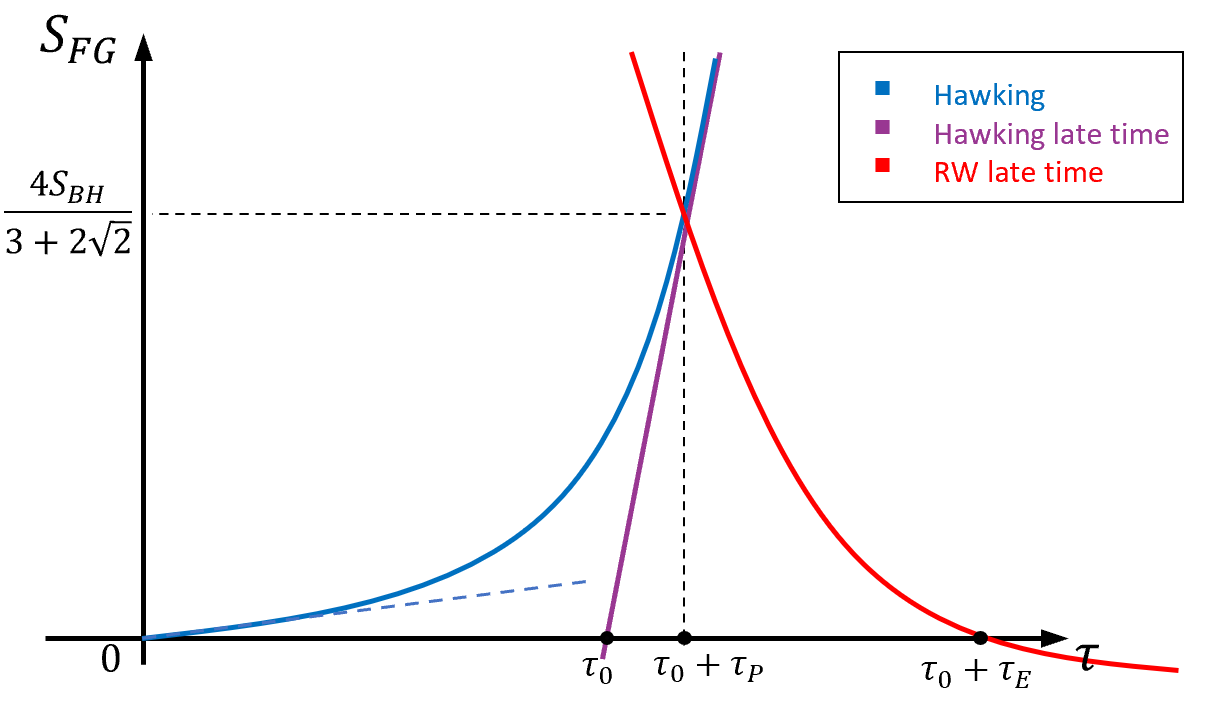}
    \caption{Page curve for the quantum-corrected collapse scenario is depicted here. Blue line represents the fine-grained entropy along the Hawking's saddle point, while the red line show the entropy along the replica-wormholes saddle point at late times. The purple line show the asymptotic behavior of the fine-grained entropy along the Hawking's saddle point at late times.}\label{sl5}
    \end{center}
\end{figure}

 Now we calculate the contribution to the fine-grained entropy from the replica-wormholes saddle point. First, let us analyze the matter term. After a careful substitution of the position of the island (\ref{delta_I_q}-\ref{x_hat_S_q}) into equation (\ref{island}) we get:
\begin{equation}
    S_{\text{matter-RW}}=\frac{1}{12}\ln{\left(\frac{4\tilde{\varepsilon}\left(-\ln{\delta_{S}^{(0)}}\right)^{2}}{-3\mathrm{F}^{+}_{\infty}(\kappa\epsilon)^{4}}\right)}-\frac{\tilde{\varepsilon}}{6}\frac{1}{1+\frac{\Delta\tau}{2\bar{a}}}e^{-\frac{\Delta\tau}{2\bar{a}}},\label{S_matter_RW}
\end{equation}
where we have used:
\begin{align}
    &\tilde{\mathrm{F}}^{+}\mathrm{F}^{-}e^{2\rho}\bigg{|}_{I}=\frac{\tilde{\varepsilon}}{4}\hspace{2mm}\land\hspace{2mm}\mathrm{F}^{-}_{I}=-\frac{\tilde{\varepsilon}}{4}\delta_{I}\hspace{2mm}\land\hspace{2mm}\tilde{\mathrm{F}}^{+}_{I}=-1,\nonumber\\
    &\mathrm{F}^{+}\mathrm{F}^{-}e^{2\rho}\bigg{|}_{S}=-1\hspace{2mm}\land\hspace{2mm}\mathrm{F}^{-}_{S}=\frac{3}{4}\tilde{\varepsilon}\delta_{I}\hspace{2mm}\land\hspace{2mm}\mathrm{F}^{+}_{S}=\mathrm{F}^{+}_{\infty},\nonumber\\
    &\text{Coord-part}=\left(-\ln{\delta_{S}^{(0)}\tilde{\varepsilon}\delta_{I}}\right)^{2}\left(1-\frac{2\tilde{\varepsilon}\delta_{I}}{1-\ln{\delta_{I}}}\right).\label{z1}
\end{align}
In (\ref{z1}), $\tilde{\mathrm{F}}^{+}_{I}$ is a coordinate transformation with respect to the $\sigma^{+}$ coordinate, while $\mathrm{F}^{+}_{S}$ is a coordinate transformation with respect to $\hat{\sigma}^{+}$. The different transformations have been used for convenience. Also, the Coord-part stands for the part of the formula (\ref{island}) under the logarithm that has direct dependence on the $\sigma$ coordinates. Since (\ref{S_matter_RW}) still depends on the UV cut-off $\epsilon$, regularization has to be done. Next, we calculate the area term in (\ref{area_term}). After a simple substitution of (\ref{x_I_q}), we get:
\begin{equation}
    \frac{A[\partial I]}{4G^{(4)}_{\text{N}}}=\frac{1}{12\tilde{\varepsilon}}\left(1-\frac{\tilde{\varepsilon}}{4}\frac{\Delta\tau}{2\bar{a}}\right)^{2}.\label{S_are_RW}
\end{equation}
The regularization selected, aligning with the observation that the island intersects the singularity close to the evaporation's end-point, is the one that ensures the fine-grained entropy vanishes. With this fact in mind as well as the minimization prescription we arrive at the following final expression for the fine-grained entropy of the Hawking radiation at late times $\tau$:
\begin{widetext}
    \begin{equation}
        \boxed{S_{FG}=\min{\left\{\frac{\Delta\tau}{24\bar{a}}+\frac{1}{6}\ln{\frac{\tau}{2b}}-\frac{1}{12}\ln{\left(\frac{-3\varepsilon\mathrm{F}^{+}_{\infty}}{4(\lambda a)^{2}}\right)},\frac{(\lambda a)^{2}}{12\varepsilon}\left(1-\frac{\varepsilon}{4(\lambda a)^{2}}\frac{\Delta\tau}{2\bar{a}}\right)^{2}-\frac{1}{12}-\frac{\varepsilon}{6(\lambda a)^{2}}\frac{1}{1+\frac{\Delta\tau}{2\bar{a}}}e^{-\frac{\Delta\tau}{2\bar{a}}}\right\}}}.\label{S_FG_q}
    \end{equation}
\end{widetext}
Note that the fine-grained entropy vanishes at the end-point of the evaporation. The page time is calculated by equating the functions within the (\ref{S_FG_q}). The result is given by:
\begin{equation}
    \kappa\tau_{P}=\frac{4(\lambda a)^{2}}{\varepsilon}\left(3-2\sqrt{2}\right).\label{t_page_q}
\end{equation}
At the Page time, the fine-grained entropy rises to its maximum value, given by:
\begin{equation}
    S_{FG}^{\text{max}}=\frac{4}{3+2\sqrt{2}}S_{BH}^{\mathrm{(class)}}.\label{S_FG_max}
\end{equation}
From (\ref{S_FG_max}) we can conclude that at every point in time, the fine-grained entropy of Hawking radiation does not cross the coarse-grained limit given by the thermodynamic entropy, as expected. In Figure \ref{sl5} the Page curve in a quantum-corrected collapse scenario is depicted.
\par
\begin{figure}
    \begin{center}
    \includegraphics[width=7cm, height=5.5cm]{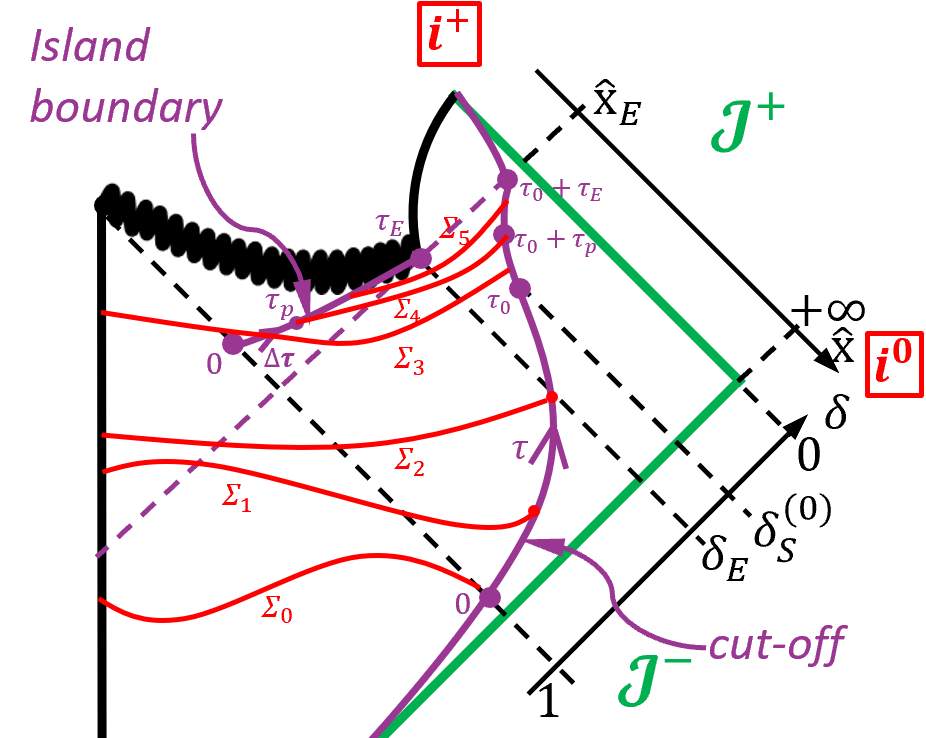}
    \caption{The complete process of black hole evaporation is shown here. Constant time slices $\Sigma_{0}-\Sigma_{3}$ correspond to the no-island case, while hypersurfaces $\Sigma_{4}$ and $\Sigma_{5}$ correspond to times after the Page time.}\label{sl6}
    \end{center}
\end{figure}

 Let us now summarize the whole process of black hole evaporation. It is illustratively shown in Figure \ref{sl6}. The purple line represents the cut-off hypersurface, parameterized by the time of evaporation $\tau$. The other purple line represents the position of the island in time, parameterized by $\Delta\tau$. Both $\tau$ and $\Delta\tau$ start from zero along the $\delta=1$ hypersurface. The red lines correspond to the region $A$ that contains the degrees of freedom inaccessible to the observer. The process will be observed from the perspective of $\tau$ time. As $\tau$ passes, the black hole starts to evaporate. At the beginning, the minimum of (\ref{S_FG_q}) is given by the Hawking's saddle point, hance the red lines end at the boundary of space-time $\mathrm{x}=0$. When $\tau$ reaches $\tau=\tau_{0}$, the island is formed within region II of space-time for the first time. This moment corresponds to $\Delta\tau=0$ along the island-boundary line. Still, the fine-grained entropy is dominated by Hawking's contribution, implying that the red lines end at the boundary. At $\tau=\tau_{0}+\tau_{P}$, or $\Delta\tau=\tau_{P}$, the phase transition occurs. From this point onward, the red lines end at the island-boundary line. The evaporation process continues until $\tau_{f}=\tau_{0}+\tau_{E}$, or when the island hits the singularity, that is, $\Delta\tau\approx\tau_{E}$. The moment $\tau=\tau_{f}$ coincides with the point at which the cut-off surface crosses into the region III of space-time, thus exiting the area where the black hole radiation exists.
\par Note that we cannot be sure of what exactly happens at the end-point of the evaporation, since at this point the distance to the end-point becomes so small that it is comparable to the Planck length. This means that quantum gravity effects take place at this moment, and the semi-classical approximation breaks. Still, we can be certain of the reproduced Page curve up its final few points.
\par 
\section{Summery and Conclusion} 
We studied the thermodynamic properties, that is, the Page curve, for the two-dimensional model of dilaton gravity derived by the reduction in dimensions of the four-dimensional Einstein-Hilbert action. Firstly we examined the classical case, where no back-reaction of the quantum fields on the geometry of space-time has been included. The result was an incorrect Page curve that saturates at the value of Bekenstein-Hawking entropy, similarly to the case of the eternal black hole \cite{DREH1}. It is shown in Figure \ref{sl4}. Then we investigated the Page curve in a case of a quantum-corrected Schwarzschild black hole solution \cite{DREH2}, which has a back-reaction included. In this case, we successfully reproduced the Page curve, depicted in Figure \ref{sl5}. The fine-grained entropy increases with time until it reaches the Page time (\ref{t_page_q}) when a phase transition occurs and the entropy starts to decrease. 
\par Within the process of obtaining the Page curve, the position of the island was calculated (\ref{delta_I_q}-\ref{x_hat_S_q}). It forms sometime after the beginning of the evaporation close to the event horizon, and then it follows the horizon until it crosses the line of singularity near the end-point of the evaporation. The whole process is illustrated in Figure \ref{sl6}.
\par The continuation of this work may include an investigation of the behavior of the Page curve for eternal and evaporating black holes when other charges are present. For example, the case of a Reissner-Nordstrom black hole may be analyzed. Although the influence of quadratic curvature terms on the Ryu-Takayanagi formula has been previously analyzed, the Page curve for a black hole with torsion remains unexplored. To address this, one could formulate and explore the dimensional reduction of five-dimensional Chern-Simons theory.

{\bf Acknowledgments}
The authors acknowledge the funding provided by the Faculty of Physics of the University of Belgrade, through a grant number 451-03-47/2023-01/200162 from the Ministry of Education, Science, and Technological Development of the Republic of Serbia. Also S.\Dj. is very grateful to Ana \Dj or\dj evi\'c (University of Belgrade) for the creation of the figures for this paper. 

\bibliography{reference}
\appendix
\setcounter{figure}{0}
\renewcommand{\thefigure}{A\arabic{figure}}
\section*{APPENDIX}
\section{Details on the quantum corrected solution}\label{app_1}
In this section we delve into the details regarding the solution (\ref{rho_resenje_>}-\ref{x_resenje_<}). For more details, see \cite{DREH2}. Firstly, we define all the functions that appear in the solution:
\begin{align}
    \mathrm{S}_{-1}(\mathrm{x})&=2\left(\frac{1}{\mathrm{x}-1}-\ln{|\mathrm{x}-1|}-\frac{7}{2}\right),\nonumber\\
    \mathrm{S}_{n}^{>}(\mathrm{x})&=\int\frac{(2\mathrm{x}-1)\de\mathrm{x}}{(\mathrm{x}-1)^{2}}\left(1-\frac{1}{\mathrm{x}}\right)^{n}\mathrm{P}_{n-1}\left(\frac{1}{\mathrm{x}}\right)\nonumber\\
    &+\int\frac{(2\mathrm{x}-3)\de\mathrm{x}}{(\mathrm{x}-1)^{2}}\left(1-\frac{1}{\mathrm{x}}\right)^{n}\mathrm{Q}_{n-1}\left(\frac{1}{\mathrm{x}}\right)\nonumber\\
    &-\left(1-\frac{1}{\mathrm{x}}\right)^{n}\frac{\mathrm{P}_{n-1}\left(\frac{1}{\mathrm{x}}\right)+3\mathrm{Q}_{n-1}\left(\frac{1}{\mathrm{x}}\right)}{\mathrm{x}-1}+c_{n},\nonumber\\
    \mathrm{S}_{n}^{<}(\mathrm{x})&=\mathrm{P}_{n-1}(1)\int\frac{(2\mathrm{x}-1)\de\mathrm{x}}{(\mathrm{x}-1)^{2}}e^{n(\mathrm{x}-1)}(1-\mathrm{x})^{n}\nonumber\\
    &+\mathrm{Q}_{n-1}(1)\int\frac{(2\mathrm{x}-3)\de\mathrm{x}}{(\mathrm{x}-1)^{2}}e^{n(\mathrm{x}-1)}(1-\mathrm{x})^{n}\nonumber\\
    &+e^{n(\mathrm{x}-1)}(1-\mathrm{x})^{n-1}(\mathrm{P}_{n-1}(1)+3\mathrm{Q}_{n-1}(1))+\Tilde{c}_{n},\label{I_clanovi}
\end{align}
\noindent The constants $c_{n}$ and $\tilde{c}_{n}$ are defined so $\mathrm{S}^{>/<}_{n}(1)=0$; while polynomials $\mathrm{P}_{n}$ and $\mathrm{Q}_{n}$ both satisfy the following recurrence relation:
\begin{equation}
    \mathrm{P}_{n}(x)=\left[n(x+1)+x^{2}\frac{\de}{\de x}\right]\mathrm{P}_{n-1}(x),\label{P_rekurentna}
\end{equation}
with the initial conditions $\mathrm{P}_{0}(x)=1$ and $\mathrm{Q_{0}}(x)=x$. The reason behind the appearance of these polynomials in the solution lies in the fact that one has to solve the transcendental equation (\ref{colaps_>_x_resenje}) in terms of $\hat{\mathrm{x}}=\hat{\mathrm{x}}(\mathrm{x},\delta)$. This solution defines polynomials $\mathrm{P}_{n}(x)$: 
\begin{equation}
    \hat{\mathrm{x}}=\begin{cases}
                        1-\sum\limits_{n=1}^{\infty}\frac{\delta^{n}}{n!}e^{n(\mathrm{x}-1)}(1-\mathrm{x})^{n}\mathrm{P}_{n-1}(1),&\mathrm{x}\leqslant1\\
                        \mathrm{x}-\sum\limits_{n=1}^{\infty}\frac{(1-\delta)^{n}}{n!}\left(1-\frac{1}{\mathrm{x}}\right)^{n}\mathrm{P}_{n-1}\left(\frac{1}{\mathrm{x}}\right),&\mathrm{x}>1
                     \end{cases},\label{x_het_od_x_delta}
\end{equation}
Also, one needs to express $\ln{\hat{\mathrm{x}}}$ in terms of $\mathrm{x}$ and $\delta$. This expansion defines polynomials $\mathrm{Q}_{n}(x)$:
\begin{equation}
    \ln{\hat{\mathrm{x}}}=\begin{cases}
                        \hspace{6.5mm}-\sum\limits_{n=1}^{\infty}\frac{\delta^{n}}{n!}e^{n(\mathrm{x}-1)}(1-\mathrm{x})^{n}\mathrm{Q}_{n-1}(1),&\mathrm{x}\leqslant1\\
                        \ln{\mathrm{x}}-\sum\limits_{n=1}^{\infty}\frac{(1-\delta)^{n}}{n!}\left(1-\frac{1}{\mathrm{x}}\right)^{n}\mathrm{Q}_{n-1}\left(\frac{1}{\mathrm{x}}\right),&\mathrm{x}>1
                     \end{cases},\label{lnx_het_od_x_delta}
\end{equation}
The function $\mathrm{S}_{0}(\mathrm{x})$ is given by:
\begin{widetext}
\begin{equation}
    \mathrm{S}_{0}(\mathrm{x})=\begin{cases}
        2\frac{\mathrm{x^{2}}-2\mathrm{x}+2}{\mathrm{x}-1}\ln{\mathrm{x}}-(2\mathrm{x}-1)\ln{(1-\mathrm{x})}-\ln^{2}(1-\mathrm{x})-2\mathrm{x}+\frac{2}{\mathrm{x}-1}&;\mathrm{x}\leqslant1\\
        -(2\mathrm{x}-1)\ln{\left(1-\frac{1}{\mathrm{x}}\right)}-\ln^{2}(\mathrm{x}-1)-2\mathrm{L}(\mathrm{x})+\frac{2}{\mathrm{x}-1}&;\mathrm{x}\geqslant1\\
    \end{cases},\label{S0_definicija}
\end{equation}
\end{widetext}
where function $\mathrm{L}(\mathrm{x})=\frac{\pi^{2}}{6}-\int_{0}^{\mathrm{x}}\frac{ds}{s-1}\ln{s}$, so that $\mathrm{L}(1)=0$. Next, the quantum-corrected coordinate transformations appearing in (\ref{rho_resenje_>}) and (\ref{rho_resenje_<}) can be expressed as:
\begin{align}
    \mathrm{F}^{-}&=\left(1-\frac{1}{\hat{\mathrm{x}}}\right)\left[1+\frac{\varepsilon}{8(\lambda a)^{2}}\frac{\hat{\mathrm{x}}-1}{\hat{\mathrm{x}}}\frac{\de\mathrm{S}_{0}(\hat{\mathrm{x}})}{\de\hat{\mathrm{x}}}\right],\label{F-_resenje_final}\\
    \mathrm{F}^{+}&=-1+\frac{\varepsilon}{8(\lambda a)^{2}}\sum_{n=1}^{\infty}\frac{(1-\delta)^{n}}{n!}(z_{n+1}-nz_{n});\label{F+_resenje_final}
\end{align}
while the new asymptotically flat coordinates are given by:
\begin{align}
    \hat{\sigma}^{-}&=\sigma^{+}_{0}-2a\left[\hat{\mathrm{x}}+\ln{|\hat{\mathrm{x}}-1|}-\frac{\varepsilon}{8(\lambda a)^{2}}\mathrm{S}_{0}(\hat{\mathrm{x}})\right],\label{sigma-_hat_resenje_final}\\
    \hat{\sigma}^{+}&=\sigma^{+}+\frac{2a\varepsilon}{8(\lambda a)^{2}}\left[\frac{\pi^{2}}{3}+2+\sum_{n=1}^{\infty}\frac{(1-\delta)^{n}}{n!}z_{n}\right].\label{sigma+_hat_resenje_final}
\end{align}
Note that $\mathrm{S}_{-1}(\mathrm{x})$ and $\mathrm{S}_{0}(\mathrm{x})$ are not well defined when $\mathrm{x}=1$ or $\mathrm{x}=0$. This represents a great problem in terms of finding the position of the apparent horizon, event horizon, singularity, or any hypersurface that appears near the $\mathrm{x}=1$ line (see \cite{DREH2}). This problem can be mended by a more careful integration of the equations of motion of the theory. For this reason, we introduce a new coordinate:
\begin{equation}
    \mathrm{y}=\frac{\mathrm{x}}{1+\frac{\tilde{\varepsilon}}{4}\ln{\delta}}\equiv\frac{\mathrm{x}}{\tilde{a}},\label{y_deff}
\end{equation}
where $\tilde{\varepsilon}=\frac{\varepsilon}{4(\lambda a)^{2}}$ and $\tilde{a}=1+\frac{\tilde{\varepsilon}}{4}\ln{\delta}$. Also, we define a new function:
\begin{equation}
    \mathcal{J}(\mathrm{y})=-\frac{4}{\tilde{\varepsilon}}+\left(\frac{4}{\tilde{\varepsilon}}+\frac{9}{2}\right)\exp{\left(\frac{\tilde{\varepsilon}}{4}\tilde{f}(\mathrm{y})\right)},\label{J_resenje}
\end{equation}
where function $\tilde{f}(\mathrm{y})$ is given by: 
\begin{widetext}
    \begin{align}
        \tilde{f}(\mathrm{z})&=\frac{4}{\tilde{\varepsilon}}\ln{\bigg{|}\frac{\mathrm{z}^{2}+\alpha_{2}\mathrm{z}+\beta_{2}}{1-\mathrm{z}^{2}}\bigg{|}}\label{f_tilde_resenje}\\
        &\hspace{5mm}+\left(1-\frac{5}{4}\tilde{\varepsilon}\right)\ln{|\mathrm{z}-\mathrm{z}_{1}^{+}|}-\left(1-\frac{1}{4}\tilde{\varepsilon}\right)\ln{|\mathrm{z}-\mathrm{z}_{1}^{-}|}-\left(1-\frac{3}{4}\tilde{\varepsilon}\right)\ln{|\mathrm{z}-\mathrm{z}_{2}^{+}|}+\left(1+\frac{3}{4}\tilde{\varepsilon}\right)\ln{|\mathrm{z}-\mathrm{z}_{2}^{-}|}+\frac{\tilde{\varepsilon}}{4}\ln{\frac{\tilde{\varepsilon}}{16}},\nonumber
    \end{align}
\end{widetext}
where $\mathrm{z}=\sqrt{\frac{\mathrm{y}-\sqrt{\tilde{\varepsilon}}}{\mathrm{y}+\sqrt{\tilde{\varepsilon}}}}$. The constants $\mathrm{z}^{\pm}_{1,2}$ that appear in equation (\ref{f_tilde_resenje}) are given by:
\begin{align}
    \mathrm{z}_{2}^{+}&=1+\frac{\tilde{\varepsilon}^{\frac{3}{2}}}{4}-\frac{\tilde{\varepsilon}^{\frac{5}{2}}}{16}+\frac{\tilde{\varepsilon}^{3}}{32}+\frac{\tilde{\varepsilon}^{\frac{7}{2}}}{32}-\frac{\tilde{\varepsilon}^{4}}{64}\label{z2+}\\
    \mathrm{z}_{2}^{-}&=-1-\frac{\tilde{\varepsilon}^{\frac{3}{2}}}{4}-\frac{\tilde{\varepsilon}^{\frac{5}{2}}}{16}-\frac{\tilde{\varepsilon}^{3}}{32}-\frac{\tilde{\varepsilon}^{\frac{7}{2}}}{32}-\frac{\tilde{\varepsilon}^{4}}{64}\label{z2-}\\
    \mathrm{z}_{1}^{+}&=1-\tilde{\varepsilon}^{\frac{1}{2}}+\frac{\tilde{\varepsilon}}{2}-\frac{\tilde{\varepsilon}^{\frac{3}{2}}}{4}+\frac{\tilde{\varepsilon}^{2}}{8}-\frac{\tilde{\varepsilon}^{\frac{5}{2}}}{16}+\frac{\tilde{\varepsilon}^{3}}{32}-\frac{\tilde{\varepsilon}^{4}}{128},\label{z1+}\\
    \mathrm{z}_{1}^{-}&=-1-\tilde{\varepsilon}^{\frac{1}{2}}-\frac{\tilde{\varepsilon}}{2}+\frac{\tilde{\varepsilon}^{\frac{3}{2}}}{4}+\frac{3\tilde{\varepsilon}^{2}}{8}+\frac{3\tilde{\varepsilon}^{\frac{5}{2}}}{16}-\frac{\tilde{\varepsilon}^{3}}{32}+\frac{5\tilde{\varepsilon}^{4}}{128},\label{z1-}
\end{align}
and also $\alpha_{2}$ and $\beta_{2}$ can be expressed by the following equations:
\begin{align}
    \alpha_{2}&=\frac{\tilde{\varepsilon}^{\frac{5}{2}}}{8}\left(1+\frac{\tilde{\varepsilon}^{\frac{3}{2}}}{4}\right),\label{alpha_2}\\
    \beta_{2}&=-1-\frac{\tilde{\varepsilon}^{\frac{3}{2}}}{2}\left(1+\frac{\tilde{\varepsilon}^{\frac{3}{2}}}{4}+\frac{\tilde{\varepsilon}^{2}}{8}\right).\label{beta_2}
\end{align}
With the help of this newly defined notation (\ref{J_resenje}-\ref{beta_2}), equations (\ref{x_resenje_>}) and (\ref{x_resenje_<}) become well defined around $\mathrm{x}=1$:
\begin{widetext}
    \begin{align}
        &\left(1+\frac{\varepsilon}{4(\lambda a)^{2}}\ln{\delta}\right)\mathcal{J}(\mathrm{y})-\frac{\varepsilon}{8(\lambda a)^{2}}\left[(2\mathrm{y}+3)\ln{\mathrm{y}}+5\mathrm{y}-2\mathrm{L}(\mathrm{y})-\sum_{n=1}^{\infty}\frac{(1-\delta)^{n}}{n!}\mathrm{S}_{n}^{>}(\mathrm{y})\right]\nonumber\\
        &\hspace{5cm}=-\ln{\delta}+\mathcal{J}(\hat{\mathrm{x}})-\frac{\varepsilon}{8(\lambda a)^{2}}\left[(2\hat{\mathrm{x}}+3)\ln{\hat{\mathrm{x}}}+5\hat{\mathrm{x}}-2\mathrm{L}(\hat{\mathrm{x}})\right]=\frac{r_{*}}{a},\hspace{4mm}\text{when }\mathrm{y}>1\label{y_>_resenje}\\
        &\left(1+\frac{\varepsilon}{4(\lambda a)^{2}}\ln{\delta}\right)\mathcal{J}(\mathrm{y})-\frac{\varepsilon}{8(\lambda a)^{2}}\left[\frac{2\mathrm{y}^{2}}{\mathrm{y}-1}\ln{\mathrm{y}}+3\mathrm{y}-\sum_{n=1}^{\infty}\frac{\delta^{n}}{n!}\mathrm{S^{<}_{n}}(\mathrm{y})\right]\nonumber\\
        &\hspace{5cm}=-\ln{\delta}+\mathcal{J}(\hat{\mathrm{x}})-\frac{\varepsilon}{8(\lambda a)^{2}}\left[(2\hat{\mathrm{x}}+3)\ln{\hat{\mathrm{x}}}+5\hat{\mathrm{x}}-2\mathrm{L}(\hat{\mathrm{x}})\right]=\frac{r_{*}}{a},\hspace{4mm}\text{when }\mathrm{y}<1.\label{y_<_resenje}
    \end{align}
\end{widetext}
Now, using equation (\ref{y_<_resenje}) it is easy to find the general solution for $\hat{\mathrm{x}}$ when $\mathrm{y}=1+\tilde{\varepsilon}\eta$:
\begin{equation}
    \hat{\mathbf{x}}=1+\frac{\tilde{\varepsilon}}{4}\bigg{(}1+(4\eta-1)\delta\bigg{)}.\label{resenje_y=1}
\end{equation}
For the calculation of the Page curve, derivatives $\partial_{\pm}\mathrm{x}$ and $\partial_{\pm}\rho$ will be needed. They are given by the following expressions:
\begin{widetext}
    \begin{align}
        \partial_{+}\mathrm{x}&=\frac{1}{2a}\left\{1-\frac{1}{\sqrt{\mathrm{y}^{2}-\tilde{\varepsilon}}}\left[1-\frac{\tilde{\varepsilon}}{8}\left(2(\mathrm{y}-2)\ln{\mathrm{y}}-8+5\mathrm{y}+\frac{3}{\mathrm{y}}-(\mathrm{y}-1)\sum_{n=1}^{\infty}\frac{\delta^{n}}{n!}\left(\frac{\mathrm{y}-1}{\mathrm{y}}\frac{\mathrm{d}\mathrm{S}_{n}^{<}}{\mathrm{dy}}-n\mathrm{S}_{n}^{<}\right)\right)\right]\right\},\label{izvod_x_+}\\
        \partial_{-}\mathrm{x}&=-\frac{1}{2a\mathrm{F}^{-}}\left\{1-\frac{1}{\sqrt{\mathrm{y}^{2}-\tilde{\varepsilon}}}\left[1-\frac{\tilde{\varepsilon}}{8}\left(2(\mathrm{y}-2)\ln{\mathrm{y}}-2(\mathrm{y}-1)\ln{(1-\mathrm{y})}+1-2\mathrm{y}+\frac{3}{\mathrm{y}}-\frac{(\mathrm{y}-1)^{2}}{\mathrm{y}}\sum_{n=1}^{\infty}\frac{\delta^{n}}{n!}\frac{\mathrm{d}\mathrm{S}_{n}^{<}}{\mathrm{dy}}\right)\right]\right\},\label{izvod_x_-}\\
        \partial_{+}\rho&=\frac{1}{4a\tilde{a}}\frac{\mathrm{y}}{(\mathrm{y}^{2}-\tilde{\varepsilon})^{\frac{3}{2}}}\left\{1+\frac{\tilde{\varepsilon}}{4}\left[(3+2\mathrm{y})\left(1-\frac{1}{\mathrm{y}^{2}}\right)+2\ln{\mathrm{y}}+\frac{1}{2}\sum_{n=1}^{\infty}\frac{\delta^{n}}{n!}\left(n(1-n\mathrm{y}^{2})\mathrm{S}_{n}^{<}-\frac{\mathrm{y}-1}{\mathrm{y}}(\mathrm{y}^{2}+\mathrm{y}-1)\frac{\mathrm{dS}_{n}^{<}}{\mathrm{dy}}\right)\right]\right\},\label{izvod_rho_+}\\
        \partial_{-}\rho&=-\frac{1}{4a\mathrm{F}^{-}}\left\{\frac{\hat{\mathrm{x}}^{2}-1}{\hat{\mathrm{x}}^{2}}\left(1-\frac{\tilde{\varepsilon}}{4}\mathrm{F}(\hat{\mathrm{x}})\right)-\frac{1}{\tilde{a}}\right.\nonumber\\
        &\hspace{5mm}\left.+\frac{1}{\tilde{a}\mathrm{y}^{2}}\left[1+\frac{\tilde{\varepsilon}}{4}\left((\mathrm{y}^{2}-1)\left(\ln{(1-\mathrm{y})}+(2\mathrm{y}-3)\frac{\mathrm{y}^{2}+2}{2\mathrm{y}^{2}}\right)+2\ln{\mathrm{y}}+\frac{1}{2}(\mathrm{y}^{2}+1)(\mathrm{y}-1)^{2}\sum_{n=1}^{\infty}\frac{\delta^{n}}{n!}\frac{\mathrm{dS}_{n}^{<}}{\mathrm{dy}}\right)\right]\right\},\label{izvod_rho_-}
    \end{align}
\end{widetext}
where the function $\mathrm{F}(\hat{\mathrm{x}})$ is defined by:
\begin{equation}
    \mathrm{F}(\hat{\mathrm{x}})=\ln{\left|1-\frac{1}{\hat{\mathrm{x}}}\right|}+\frac{1}{\hat{\mathrm{x}}}-\frac{3}{2}\frac{1}{\hat{\mathrm{x}}^{2}}.\label{F_x_hat}
\end{equation}
In terms of this function, the coordinate transformation $\mathrm{F}^{-}$ can be expressed as:
\begin{equation}
    \mathrm{F}^{-}=\frac{\hat{\mathrm{x}}-1}{\hat{\mathrm{x}}}\left[1-\frac{\tilde{\varepsilon}}{4}\left(\mathrm{F}(\hat{\mathrm{x}})+\frac{2\hat{\mathrm{x}}-1}{\hat{\mathrm{x}}^{2}(\hat{\mathrm{x}}-1)}\right)\right].\label{F_-_od_F_x_hat}
\end{equation}
Notice that all expressions (\ref{y_>_resenje}-\ref{izvod_rho_-}) are well defined around the $\mathrm{y}=1$ and $\hat{\mathrm{x}}=1$ hypersurfaces. At the end of this section we present the expressions for the apparent horizon, event horizon and the singularity in terms of $\mathrm{y}=\mathrm{y}(\delta)$:
\begin{align}
    \mathrm{y}_{AH}(\delta)&=\sqrt{1+\tilde{\varepsilon}}\label{y_AH}\\
    \mathrm{y}_{H}(\delta)&=1+\frac{\tilde{\varepsilon}}{4}\left(1+\frac{\delta_{E}}{\delta}\right)\label{y_H}\\
    \mathrm{y}_{S}&=\frac{\sqrt{\tilde{\varepsilon}}}{1+\frac{\tilde{\varepsilon}}{4}\ln{}\delta}\label{y_S}.
\end{align}
\section{Entanglement entropy formula}\label{app_2}
\begin{figure}
    \begin{center}
    \includegraphics[width=5cm, height=5cm]{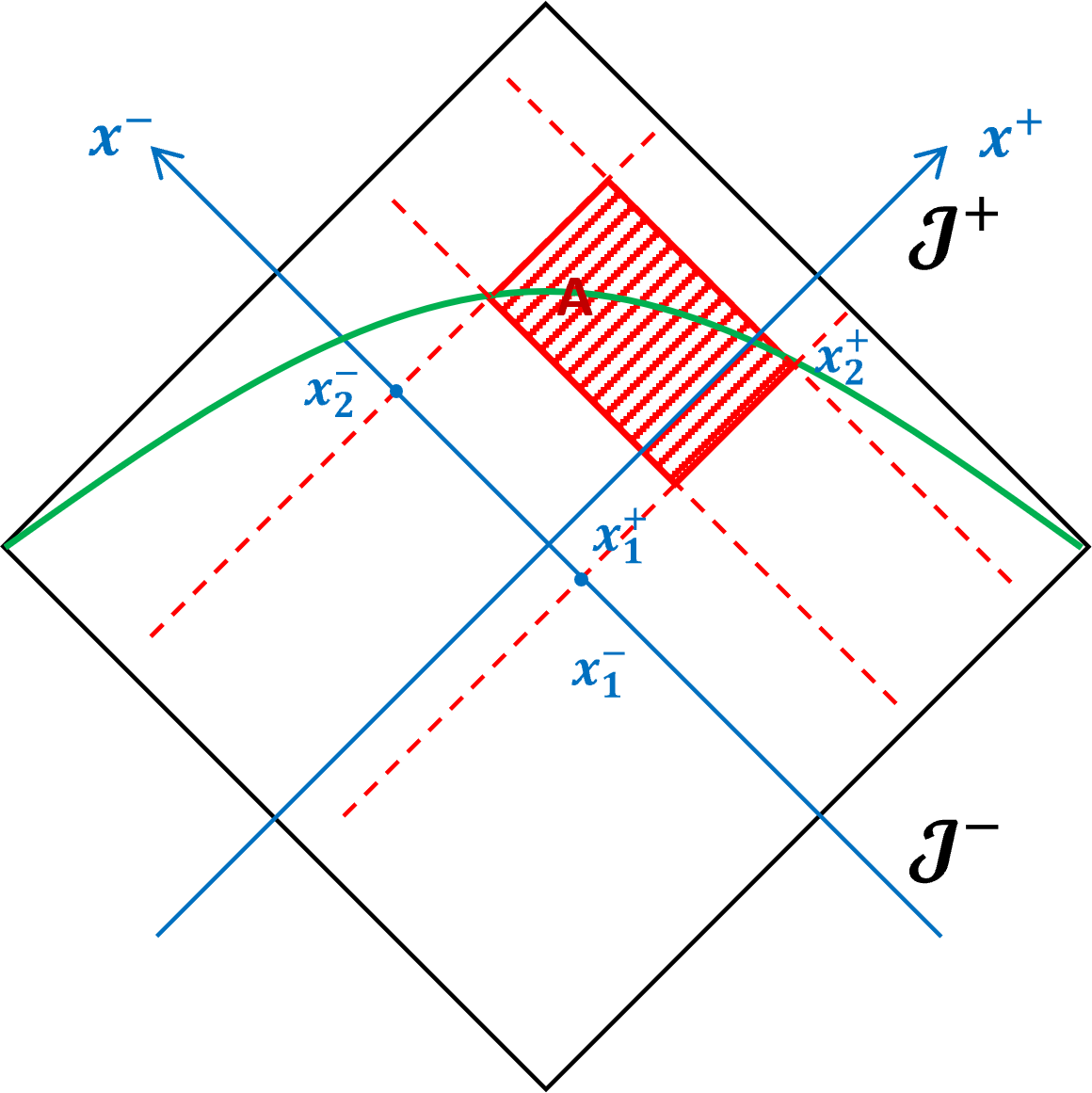}
    \caption{Position of the region ($A$) in witch the inaccessible degrees of freedom live.}\label{ent1}
    \end{center}
\end{figure}

In this section we present a brief look at the derivation of the QFT entanglement entropy formula ($S_{\mathrm{matter}}$ in equation (\ref{Island_formula})), closely following \cite{QCGHS}. Let us assume that the state of the quantum fields $\ket{\psi}$, which comprise matter, is pure within the whole space-time; and that there exists an observer that does not have access to the degrees of freedom that live in a portion $A$ of space-time (see Figure \ref{ent1}). In addition, let us assume that the matter is given by the massless scalar field; then the equation of motion for this field is $\partial_{+}\partial_{-}f=0$. The general solution to this equation can be expressed as $f(x^{+},x^{-})=f(x^{+})+f(x^{-})$, where $f(x^{\pm})$ represent the left-propagating and the right-propagating modes of the field. Since they are distinct, one can analyze the entropy of each type of modes separately and then add them together to get the entropy of the whole field. To obtain the density matrix of radiation outside the region $A$, one needs to trace out the degrees of freedom living inside the region $A$, $\hat{\rho}_{out}=\mathrm{Tr}_{in}\{\ket{\psi}\bra{\psi}\}$. Then, the entanglement entropy is calculated according to the following von Neumann formula $S_{ent}=-k_{B}\mathrm{Tr}\{\hat{\rho}_{out}\ln{\hat{\rho}_{out}}\}$.
\par Let us first recall Unruh's result for the entropy of a uniformly accelerating observer in Minkowski space-time, also known as the Rindler observer. If the world line of a Rindler observer belongs to the right Rindler wedge, the observer cannot access the degrees of freedom that live in the causally disconnected left Rindler wedge (region $A$). Taking into account that the vacuum state is that of the Minkowski vacuum, after a detailed calculation for the entropy of the right-propagating modes, one gets:
\begin{equation}
S_{ent}=\frac{1}{12}\ln{\frac{X^{-}_{max}}{\epsilon^{-}}},\label{E1}
\end{equation}
where $X^{-}_{max}$ stands for the IC cut-off in the light cone directions while $\epsilon^{-}$ represents a UV cut-off. To get the full entropy, we need to add the left-propagating modes; then the result is:
\begin{equation}
    S_{ent}=\frac{1}{12}\ln{\frac{X^{+}_{max}X^{-}_{max}}{\epsilon^{2}}},\label{E2}
\end{equation}
where $\epsilon=\sqrt{\epsilon^{+}\epsilon^{-}}$ is a Lorentz invariant quantity. The next step is to generalize the previous formula to include
an arbitrary inaccessible region of Minkowski space-time, see Figure \ref{ent1}. In the case of the left-propagating modes, generalization of (\ref{E1}) leads us to:
\begin{equation}
    S_{ent}=\frac{1}{12}\ln{\frac{(x^{-}_{2}-x^{-}_{1})^{2}}{\epsilon^{-}_{1}\epsilon^{-}_{2}}},\label{E3}
\end{equation}
where $\epsilon^{-}_{1,2}$ are the UV cut-offs at the left and right boundaries of the region $A$ (see Figure \ref{ent1}). After adding the left-propagating modes, the generalization of (\ref{E2}) is given by:
\begin{equation}
    S_{ent}=\frac{1}{12}\ln{\frac{(x^{+}_{2}-x^{+}_{1})^{2}(x^{-}_{2}-x^{-}_{1})^{2}}{\epsilon^{4}}}.\label{E4}
\end{equation}
For the purpose of the gravitational entanglement entropy formula, we need to generalize all previous equations to the curved space-time. The first step is to find how formula (\ref{E3}) looks in some other flat space-time coordinates $y^{\pm}=y^{\pm}(x^{\pm})$. Since the calculation is the same as in Minkowski coordinates, one gets
\begin{equation}
    S_{ent}=\frac{1}{12}\ln{\frac{(y^{-}_{2}-y^{-}_{1})^{2}}{\hat{\epsilon}^{-}_{1}\hat{\epsilon}^{-}_{2}}},\label{E5}
\end{equation}
where $\hat{\epsilon}^{-}_{1,2}$ are the UV cut-offs in $y$-coordinates at the boundaries of the region $A$. Since these cut-offs vary from one point of space-time to another, they are not proper cut-offs. They transform as length, which means that we can transform them back to Minkowski coordinates, which are globally flat. In conformal gauge, $\mathrm{d}s^{2}=-e^{2\rho}\mathrm{d}y^{+}\mathrm{d}y^{-}$, this yields:
\begin{equation}
    S_{ent}=\frac{1}{12}\ln{\frac{(y^{-}_{2}-y^{-}_{1})^{2}}{y^{-}_{1}\text{'}y^{-}_{2}\text{'}\epsilon^{-}_{1}\epsilon^{-}_{2}}}.\label{E6}
\end{equation}
After adding the right-propagating modes, the result is given by:
\begin{equation}
S_{ent}=\frac{1}{12}\ln{\frac{(y^{+}_{2}-y^{+}_{1})^{2}(y^{-}_{2}-y^{-}_{1})^{2}}{\epsilon^{4}e^{-2\rho_{1}}e^{-2\rho_{2}}}},\label{E7}
\end{equation}
where we have used $y^{-}\text{'}y^{+}\text{'}=e^{-2\rho}$. Since one can choose to define $\epsilon$ in locally flat coordinates, it is easy to conclude that formula (\ref{E7}) holds in curved space-time as well. 
\begin{figure}
    \begin{center}
    \includegraphics[width=3.5cm, height=5.3cm]{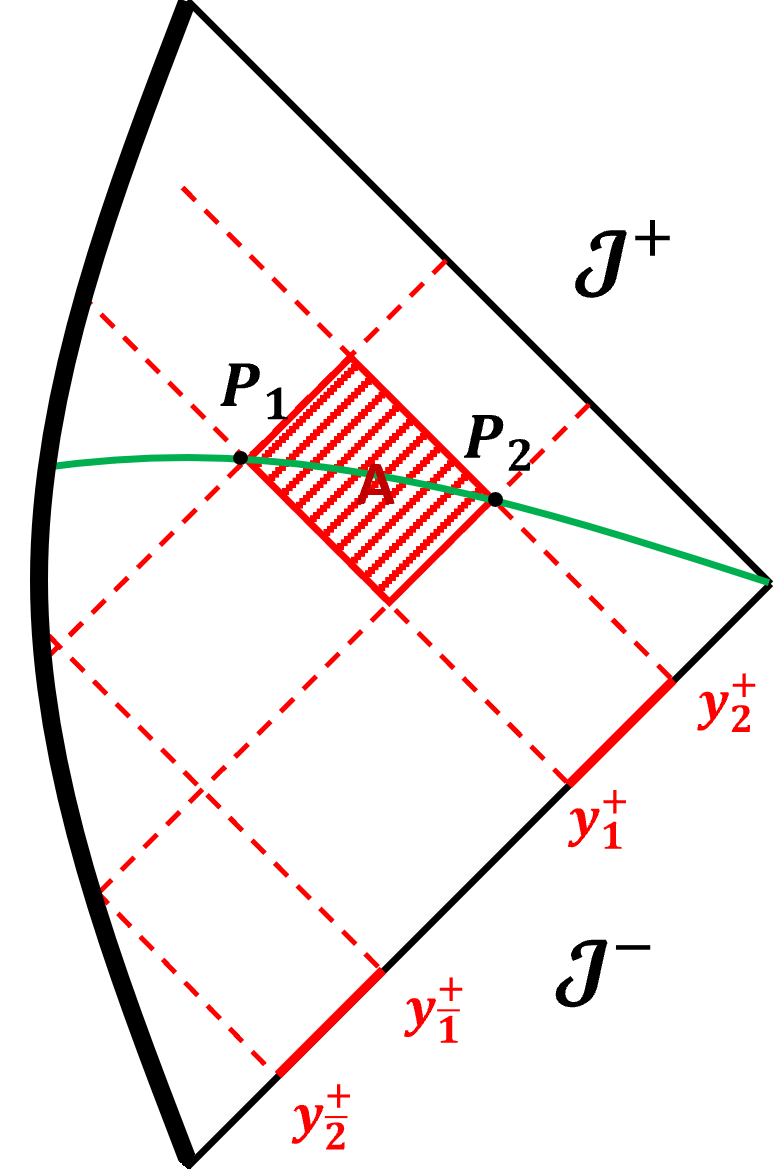}
    \caption{Position of region ($A$) in witch the inaccessible degrees of freedom live.}\label{ent2}
    \end{center}
\end{figure}
\par In the case of an evaporating black hole, there exists a boundary of space-time at the $r=0$ value of the radial coordinate. This implies that the left-propagating and right-propagating modes are no longer distinct and that they have some type of correlation. To include these correlations, we impose reflective boundary conditions at the boundary of space-time. Let us first examine the case where the region $A$ touches the boundary at one point $M$ (see Figure \ref{ent2}). It is easy to see that the right-propagating modes living in the interval $[y_{P}^{-},y_{M}^{-}]$ are equivalent to the left-propagating modes living in the interval $[y_{\bar{P}}^{+},y_{\bar{M}}^{+}]$, where coordinates $y_{\bar{P}}^{+}$ and $y_{\bar{M}}^{+}$ are determined using the definition of reflective boundary conditions at the $r=0$ boundary of space-time. The case depicted in Figure (\ref{ent2}) represents the boundary conditions, which implies: $y_{\bar{M}}^{+}=y_{M}^{-}$. Taking all this into account, the calculation of the entanglement entropy in this case reduces to the calculation of the entanglement entropy for the left-propagating modes along the interval $[y_{\bar{P}}^{+},y_{P}^{+}]$. The result is given by:
\begin{equation}
    S_{ent}=\frac{1}{12}\ln{\frac{(y_{P}^{+}-y_{\bar{P}}^{+})^{2}}{4\epsilon^{2}e^{-2\rho_{P}}}}.\label{ent_no_island}
\end{equation}
Notice that in (\ref{ent_no_island}) the UV cut-off is still given as $\epsilon=\sqrt{\epsilon^{+}\epsilon^{-}}$. The reason behind this lies in the fact that $y_{\bar{P}}^{+}$ is a mirror point of point $y^{-}_{P}$. Also, factor 4 appears due to the mirroring process.
\par
\begin{figure}
    \begin{center}
    \includegraphics[width=5cm, height=6.5cm]{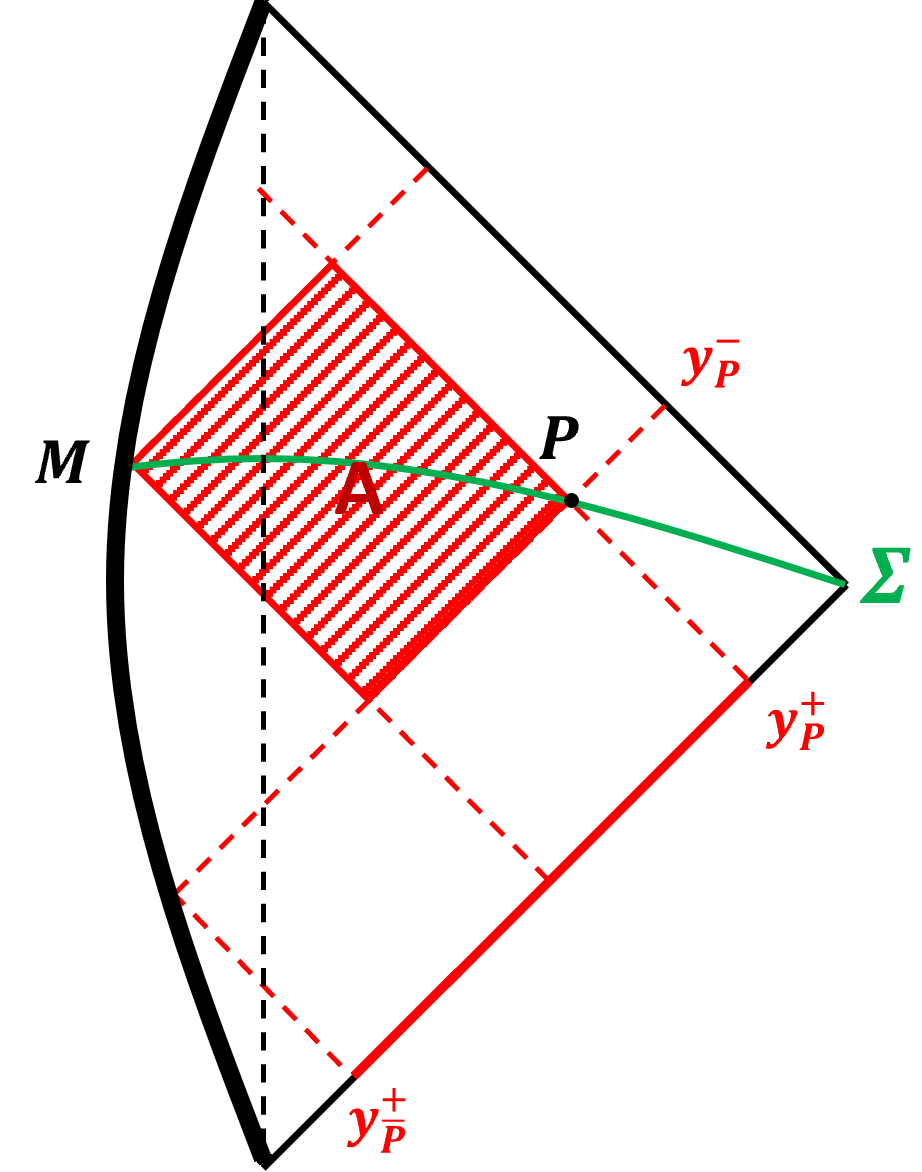}
    \caption{Position of region ($A$) in witch the inaccessible degrees of freedom live.}\label{ent3}
    \end{center}
\end{figure}

 The formula (\ref{ent_no_island}) is applicable when there is no island present. But, when the island appears, the end-point of the region $A$ moves away from the boundary, and the problem reduces to the calculation of the entanglement entropy of two disjoint regions along the $\mathcal{J}^{-}$ hypersurface, as can be seen in Figure \ref{ent3}. Using a reasoning similar to that in \cite{QCGHS}, one can obtain the following formula for entanglement entropy:
\begin{widetext}
    \begin{equation}
        S_{ent}=\frac{1}{12}\ln{\frac{(y^{+}_{1}-y^{+}_{2})^{2}(y^{+}_{\bar{1}}-y^{+}_{\bar{2}})^{2}(y^{+}_{\bar{1}}-y^{+}_{1})^{2}(y^{+}_{\bar{2}}-y^{+}_{2})^{2}}{\epsilon^{4}(y^{+}_{\bar{2}}-y^{+}_{1})^{2}(y^{+}_{\bar{1}}-y^{+}_{2})^{2}e^{-2\rho_{1}}e^{-2\rho_{2}}}}\label{ent_island}
    \end{equation}
\end{widetext}
Note that formula (\ref{ent3}) holds in curved spacetime as well.

\end{document}